%% file: arxiv_main.tex
\documentclass[letterpaper,twocolumn,10pt]{article}
\usepackage{usenix}

\usepackage{tikz}
\usepackage{amsmath}
\AtBeginDocument{%
  \providecommand\BibTeX{{%
    Bib\TeX}}}

\usepackage[most]{tcolorbox} 
\usepackage{algorithmic}
\usepackage[binary-units=true]{siunitx}
\usepackage{graphicx}
\usepackage{glossaries}
\usepackage{textcomp}
\usepackage[dvipsnames]{xcolor} 
\usepackage{hyperref}
\usepackage{cleveref}
\usepackage{booktabs}
\usepackage[shortlabels]{enumitem}
\usepackage{makecell}
\usepackage{subcaption}
\usepackage{multirow}
\usepackage{tikz}
\usepackage{amsfonts}
\usepackage[english]{babel}
\usepackage[T1]{fontenc}
\usepackage{xspace}

\def\BibTeX{{\rm B\kern-.05em{\sc i\kern-.025em b}\kern-.08em
    T\kern-.1667em\lower.7ex\hbox{E}\kern-.125emX}}

\DeclareSIUnit[per-mode=symbol]\bps{\bit\per\second}
\DeclareSIUnit[per-mode=symbol]\kbps{\kilo\bps}
\DeclareSIUnit[per-mode=symbol]\Mbps{\mega\bps}
\DeclareSIUnit[per-mode=symbol]\Gbps{\giga\bps}
\DeclareSIUnit[per-mode=symbol]\nanosec{\nano\second}
\DeclareSIUnit[per-mode=symbol]\packet{packet}
\DeclareSIUnit[per-mode=symbol]\packetps{\packet\per\second}
\DeclareSIUnit\microsec{\SIUnitSymbolMicro s}
\DeclareSIUnit\byte{B}
\DeclareSIUnit\bit{bit}
\DeclareSIUnit\terabyte{TB}

\makeglossaries

\input{ideal_functionality_macros}

\begin{document}
\glsdisablehyper

\date{}

\title{\Large \bf Proof of Cloud: Data Center Execution Assurance for Confidential VMs}

\author{
{\rm Filip Rezabek}\\
Flashbots /  Technical University of Munich
\and
{\rm Moe Mahhouk}\\
Flashbots
\and
{\rm Andrew Miller}\\
Flashbots
\and
{\rm Quintus Kilbourn}\\
Flashbots
\and
{\rm Georg Carle}\\
Technical University of Munich
\and
{\rm Jonathan Passerat-Palmbach}\\
Flashbots / Imperial College London
} 

\maketitle

\begin{abstract}
\input{abstract}
\end{abstract}

\input{acronyms}

\input{include/introduction}
\input{include/background}

\input{include/analysis}

\input{include/design}
\input{include/security_analysis}
\input{include/formal_proof}
\input{include/protocol_mitigations}
\input{include/discussion}

\input{include/conclusion}



\cleardoublepage
\appendix

\bibliographystyle{plain}
\bibliography{sample.bib}

\input{include/implementation-appendix_arxiv}

\input{include/appendix_dcea_proof}

%

\printglossary

\cleardoublepage

\end{document}

%% file: ideal_functionality_macros.tex
\definecolor{bb}{RGB}{0,0,0}
\definecolor{gg}{RGB}{220,220,220}
\definecolor{YellowOrange}{rgb}{0.98, 0.6, 0.01}
\definecolor{Red}{rgb}{0.89, 0.0, 0.13}
\definecolor{Black}{rgb}{0.0, 0.0, 0.0}
\definecolor{Purple}{rgb}{0.63, 0.36, 0.94}
\definecolor{purp}{rgb}{0.59, 0.48, 0.71}

\makeatletter
\newcounter{bbox} 

\newtcolorbox[use counter=bbox]{bbox}[4][]{%
    colback=white,
    colframe=bb,
    colbacktitle=white!40!gg,
    coltitle=black,
    fonttitle=\footnotesize\bfseries,
    fontupper=\footnotesize,
    fontlower=\footnotesize,
    enhanced,
    breakable,
    before skip=8pt plus 2pt minus 2pt,
    after skip=8pt plus 2pt minus 2pt,
    attach boxed title to top left={yshift=-2mm, xshift=0.5cm},%
    title={#3 \thebbox: #4}, 
    label={#2}, 
    #1,
}

\newcommand{\refbbox}[1]{\hyperref[#1]{Box~\ref{#1}}} 
\makeatother

\newcounter{protocolbox} 
\renewcommand{\theprotocolbox}{\arabic{protocolbox}} 

\newtcolorbox[use counter=protocolbox]{protocolbox}[3][]{%
    colback=white,
    colframe=bb,
    colbacktitle=white!40!gg,
    coltitle=black,
    fonttitle=\footnotesize\bfseries,
    fontupper=\footnotesize,
    fontlower=\footnotesize,
    enhanced,
    breakable,
    before skip=8pt plus 2pt minus 2pt,
    after skip=8pt plus 2pt minus 2pt,
    attach boxed title to top left={yshift=-2mm, xshift=0.5cm},%
    title={Protocol \theprotocolbox: #3}, 
    label={#2}, 
    #1,
}

\usepackage{xkeyval} 
\usepackage{etoolbox} 

\newcounter{functionalitybox} 

\makeatletter
\define@key{functionalitykeys}{label}{\def\func@label{#1}}
\define@key{functionalitykeys}{title}{\def\func@title{#1}}

\renewcommand{\thefunctionalitybox}{\arabic{functionalitybox}} 

\newenvironment{functionality}[1][]{%
    \setkeys{functionalitykeys}{label=, title=, #1}%
    \refstepcounter{functionalitybox}%
    \protected@edef\@currentlabel{\thefunctionalitybox}%
    \begin{bbox}[breakable]{func@temp@\arabic{functionalitybox}}{Functionality}{\func@title}%
    \ifx\func@label\@empty\else\label{\func@label}\fi%
}{%
    \end{bbox}%
}
\makeatother

\newcounter{shellbox} 

\makeatletter
\define@key{shellkeys}{label}{\def\shell@label{#1}}
\define@key{shellkeys}{title}{\def\shell@title{#1}}

\renewcommand{\theshellbox}{\arabic{shellbox}} 

\newenvironment{shell}[1][]{%
    \setkeys{shellkeys}{label=, title=, #1}%
    \refstepcounter{shellbox}%
    \protected@edef\@currentlabel{\theshellbox}%
    \begin{bbox}[breakable]{shell@temp@\arabic{shellbox}}{Shell}{\shell@title}%
    \ifx\shell@label\@empty\else\label{\shell@label}\fi%
}{%
    \end{bbox}%
}
\makeatother

\newcommand{\msf}[1]{\ensuremath{{\mathsf {#1}}}}

\newcommand{\Send}{\textbf{Send}~}
\newcommand{\Output}{\emph{Output}~}

\newcommand{\If}{\textbf{If}~}
\newcommand{\Else}{\textbf{Else}~}

\newcommand{\Init}{{\bf \color{NavyBlue} Init}~}
\newcommand{\OnInput}{{\bf \textcolor{Black} On input}~}

\newcommand{\heading}[1]{\textbf{#1}}

\makeatletter
\newcommand{\inmsg}[1]{%
(#1\checknextarg}
\newcommand{\checknextarg}{\@ifnextchar\bgroup{\gobblenextarg}{)~}}
\newcommand{\gobblenextarg}[1]{, #1\@ifnextchar\bgroup{\gobblenextarg}{)~}}
\makeatother

\usepackage{amsthm}
\newtheorem{theorem}{Theorem}

\usepackage{enumitem}
\newenvironment{itemizeless}
  {\begin{itemize}[label={}, leftmargin=*, nosep]}
  {\end{itemize}}

%% file: abstract.tex
Confidential Virtual Machines (CVMs) protect data in use by running workloads within hardware-enforced Trusted Execution Environments (TEEs). However, existing CVM attestation mechanisms only certify what code is running, not where it is running. Commercial TEEs mitigate passive physical attacks through memory encryption but explicitly exclude active hardware tampering (memory interposers, physical side channels, ...). Yet current attestations provide no cryptographic evidence that a CVM executes on hardware residing within a trusted data center where such attacks would not take place. This gap enables proxy attacks in which valid attestations are combined across machines to falsely attest trusted execution.

To bridge this gap, we introduce Data Center Execution Assurance (DCEA), a design that generates a cryptographic Proof of Cloud by binding CVM attestation to platform-level Trusted Platform Module (TPM) evidence. DCEA combines two independent roots of trust. First, the TEE manufacturer, and second, the infrastructure provider, by cross-linking runtime TEE measurements with the vTPM-measured boot CVM state. This binding ensures that CVM execution, vTPM quotes, and platform provenance all originate from the same physical chassis.

We formalize the environment's provenance and show that DCEA prevents advanced relay attacks, including a novel mix-and-match proxy attack. Using the AGATE framework in the Universal Composability model, we prove that DCEA emulates an ideal location-aware TEE even under a malicious host software stack. We implement DCEA on Google Cloud bare-metal Intel TDX instances using Intel TXT and evaluate its performance, demonstrating practical overheads and deployability. DCEA refines the CVM threat model and enables verifiable execution-location guarantees for privacy-sensitive workloads.

%% file: acronyms.tex
\newacronym{nic}{NIC}{Network Interface Card}
\newacronym{iiot}{IIoT}{Industrial Internet of Things}
\newacronym{iot}{IoT}{Internet of Things}
\newacronym{cots}{COTS}{Commercial off-the-Shelf}
\newacronym{rtt}{RTT}{Round Trip Time}
\newacronym{e2e}{E2E}{End-to-End}
\newacronym{p2p}{P2P}{Peer-to-Peer}
\newacronym{gptp}{gPTP}{generic Precision Time Protocol}
\newacronym{phc}{PHC}{PTP Hardware Clock}
\newacronym{gm}{GM}{Grandmaster Clock}
\newacronym{tc}{tc}{traffic control}
\newacronym[plural=TCLs,firstplural=traffic classes (TCLs)]{tcl}{TCL}{Traffic Class}
\newacronym{qos}{QoS}{Quality of Service}
\newacronym{ecdf}{ECDF}{Empirical Cumulative Distribution Function}
\newacronym{be}{BE}{Best Effort}
\newacronym{kpi}{KPI}{Key Performance Indicator}
\newacronym{skb}{SKB}{Socket Buffer}
\newacronym{sut}{SUT}{System Under Test}
\newacronym{phy}{PHY}{Physical Layer}
\newacronym{udp}{UDP}{User Datagram Protocol}
\newacronym{bmca}{BMCA}{Best Master Clock Algorithm}
\newacronym{tcp}{TCP}{Transmission Control Protocol}
\newacronym{os}{OS}{Operating System}
\newacronym{irq}{IRQ}{Interrupt Request}
\newacronym{cpu}{CPU}{Central Processing Unit}
\newacronym{smp}{SMP}{Symmetrical Multiprocessing}
\newacronym{smt}{SMT}{Simultaneous Multi-Threading}
\newacronym{rt}{RT}{Real-Time}
\newacronym{hw}{HW}{hardware}
\newacronym{sw}{SW}{software}
\newacronym{kc}{KC}{Key Contribution}
\newacronym{pcap}{PCAP}{Packet Capture}
\newacronym{utc}{UTC}{Coordinated Universal Time}
\newacronym{tai}{TAI}{International Atomic Time}
\newacronym{txtime}{TxTime}{transmission time}
\newacronym{macsec}{MACsec}{Media Access Control Security}
\newacronym{hpc}{HPC}{High Performance Computer}
\newacronym{lpc}{LPC}{Low Performance Computer}
\newacronym{engine}{EnGINE}{Environment for Generic In-vehicular Networking Experiments}
\newacronym{sc}{SC}{Secure Channel}
\newacronym{fpga}{FPGA}{Field Programmable Gate Array}
\newacronym{fqcodel}{FQ\_CoDel}{Fair Queuing with Controlled Delay}
\newacronym{mac}{MAC}{Media Access Control}
\newacronym{ip}{IP}{Internet Protocol}
\newacronym{ws}{WS}{window size}
\newacronym{ram}{RAM}{Random-Access Memory}
\newacronym{is}{IS}{Interframe Spacing}
\newacronym{god}{GOD}{Guaranteed Output Delivery}
\newacronym{pp}{PP}{Payment Processor}
\newacronym{dlp}{DLP}{Discrete Logarithm Problem}
\newacronym{saas}{SaaS}{Software-as-a-Service}
\newacronym{dsa}{DSA}{Digital Signature Algorithm}
\newacronym{ecdsa}{ECDSA}{Elliptic Curve Digital Signature Algorithm}
\newacronym{eddsa}{EdDSA}{Edwards-curve DSA}
\newacronym{dkg}{DKG}{Distributed Key Generation}
\newacronym{zkp}{ZKP}{Zero-Knowledge Proof}
\newacronym{PoC}{PoC}{proof-of-concept}
\newacronym{mpc}{MPC}{Multiparty Computation}
\newacronym{ot}{OT}{Oblivious Transfer}
\newacronym{vCPUs}{vCPUs}{virtual CPUs}
\newacronym{tps}{TPS}{Transactions Per Second}
\newacronym{tee}{TEE}{Trusted Execution Environment}
\newacronym{tpm}{TPM}{Trusted Platform Module}
\newacronym{txt}{TXT}{Trusted Execution Technology}
\newacronym{vm}{VM}{Virtual Machine}
\newacronym{tdx}{TDX}{Trust Domain Extensions}
\newacronym{sev}{SEV}{Secure Encrypted Virtualization}
\newacronym{snp}{SNP}{Secure Nested Paging}
\newacronym{sgx}{SGX}{Software Guard Extensions}
\newacronym{qemu}{QEMU}{Quick Emulator}
\newacronym{kvm}{KVM}{Kernel-based Virtual Machine}
\newacronym{tsn}{TSN}{Time Sensitive Networking}
\newacronym{methoda}{METHODA}{Multilayer Environment and Toolchain for Holistic NetwOrk Design and Analysis}
\newacronym{pos}{pos}{plain orchestrating service}
\newacronym{tcb}{TCB}{Trusted Computing Base}
\newacronym{dma}{DMA}{Direct Memory Access}
\newacronym{td}{TD}{Trust Domain}
\newacronym{seam}{SEAM}{Secure Arbitration Mode}
\newacronym{maccode}{MAC}{Message Authentication Code}
\newacronym{gcp}{GCP}{Google Cloud Platform}
\newacronym{sme}{SME}{AMD Secure Memory Encryption}
\newacronym{asp}{ASP}{AMD Secure Processor}
\newacronym{psp}{PSP}{AMD Platform Security Processor}
\newacronym{es}{ES}{Encrypted State}
\newacronym{sota}{SotA}{State of the Art}
\newacronym{poc}{PoC}{Proof of Concept}
\newacronym{vlek}{VLEK}{Verified Launch Enclave Key}
\newacronym{vcek}{VCEK}{Verified Chip Endorsement Key}
\newacronym{vmrk}{VMRK}{VM Root Key}
\newacronym{kds}{KDS}{Key Distribution Server}
\newacronym{ark}{ARK}{AMD Root Key}
\newacronym{ask}{ASK}{AMD SEV Key}
\newacronym{qgs}{QGS}{quote generation service}
\newacronym{pccs}{PCCS}{Provisioning Certification Caching Service}
\newacronym{pcs}{PCS}{Intel Provisioning Certification Service}
\newacronym{mpa}{MPA}{Multi-package Registration Agent}
\newacronym{mktme}{MKTME}{Multi-key Total Memory Encryption}
\newacronym{vmm}{VMM}{VM Manager}
\newacronym{tls}{TLS}{Transport Layer Security}
\newacronym{pek}{PEK}{Platform Endorsement Key}
\newacronym{csr}{CSR}{Certificate Signing Request}
\newacronym{pckcert}{PCKC}{Provisioning Certification Key Certificate}
\newacronym{pce}{PCE}{Provisioning Certificate Enclave}
\newacronym{pck}{PCK}{Provisioning Certification Key}
\newacronym{qsk}{QSK}{Quote Signing Key}
\newacronym{tdqe}{TDQE}{TD Quoting Enclave}
\newacronym{qe}{QE}{TD Quoting Enclave}
\newacronym{ak}{AK}{Attestation Key}
\newacronym{svn}{SVN}{Security Version Number}
\newacronym{r3aal}{R3AAL}{Ring3 Attestation Abstraction Library}
\newacronym{tdqd}{TDQD}{TD Quote Driver}
\newacronym{qgl}{QGL}{Quote Generation Library}
\newacronym{qvl}{QVL}{Quote Verification Library}
\newacronym{qve}{QVE}{Quote Verification Enclave}
\newacronym{crl}{CRL}{Certificate Revocation List}
\newacronym{spd}{SPD}{Seriel Presence Detect}
\newacronym{rmp}{RMP}{Reverse Map Table}
\newacronym{xts}{XTS}{XEX-based Tweaked CodeBook Mode with Ciphertext Stealing}
\newacronym{xex}{XEX}{XOR-Encrypt-XOR}
\newacronym{aes}{AES}{Advanced Encryption Standard}
\newacronym{zk}{ZK}{Zero-knowledge}
\newacronym{ecc}{ECC}{Elliptic Curve Cryptography}
\newacronym{rsa}{RSA}{Rivest–Shamir–Adleman}
\newacronym{nist}{NIST}{National Institute of Standards and Technology}
\newacronym{bls}{BLS}{Boneh-Lynn-Shacham}
\newacronym{smi}{SMI}{System Management Interrupt}
\newacronym{smm}{SMM}{System Management Mode}
\newacronym{vmpl}{VMPL}{Virtual Machine Protection Level}
\newacronym{mmio}{MMIO}{Memory-mapped I/O}
\newacronym{svsm}{SVSM}{Secure VM Service Module}
\newacronym{vtpm}{vTPM}{virtual TPM}
\newacronym{dtpm}{dTPM}{discrete TPM}
\newacronym{vmx}{VMX}{Virtual Machines Extension}
\newacronym{cvm}{CVM}{Confidential VM}
\newacronym{tcg}{TCG}{Trusted Computing Group}
\newacronym{rtmr}{RTMR}{Runtime Extendable Measurement Register}
\newacronym{mr}{MR}{Measurement Register}
\newacronym{mrtd}{MRTD}{Measurement of Trust Domain}
\newacronym{ppid}{PPID}{Protected Platform Identifier}
\newacronym{uuid}{UUID}{Universally Unique Identifier}
\newacronym{pki}{PKI}{Public Key Infrastructure}
\newacronym{as}{AS}{Autonomous System}
\newacronym{cca}{CCA}{Confidential Compute Architecture}
\newacronym{vtl}{VTL}{Virtual Trust Level}
\newacronym{dcap}{DCAP}{Data Center Attestation Primitives}
\newacronym{mbr}{MBR}{Master Boot Record}
\newacronym{pcr}{PCR}{Platform Configuration Register}
\newacronym{vt}{VT}{Virtualization Technology}
\newacronym{crtm}{CRTM}{Core Root of Trust for Measurement}
\newacronym{ek}{EK}{Endorsement Key}
\newacronym{ekc}{EKC}{EK Certificate}
\newacronym{drtm}{DRTM}{Dynamic Root of Trust Measurement}
\newacronym{mle}{MLE}{Measured Launch Environment}
\newacronym{acm}{ACM}{Authenticated Code Module}
\newacronym{dcea}{DCEA}{Data Center Execution Assurance}
\newacronym{defi}{DeFi}{Decentralized Finance}
\newacronym{aws}{AWS}{Amazon Web Services}
\newacronym{svm}{SVM}{Secure Virtual Machine}
\newacronym{tdvf}{TDVF}{Trust Domain firmware}
\newacronym{ct}{CT}{Certificate Transparency}
\newacronym{poe}{POE}{Platform Ownership Endorsement}
\newacronym{guc}{GUC}{Global Universal Composition}
\newacronym{iti}{ITI}{Turing Machine Instance}
\newacronym{ca}{CA}{Certificate Authority}
\newacronym{uc}{UC}{Universal Composition}
\newacronym{piid}{PIID}{Platform Instance Identity}
\newacronym{rim}{RIM}{Realm Initial Measurement}
\newacronym{rem}{REM}{Realm Extensible Measurement}

\newcommand{\encircled}[2][0.8mm]{%
    \raisebox{.5pt}{%
        \textcircled{%
            \raisebox{0.35pt}{%
                \kern #1
                \scalebox{0.70}{#2}
            }%
        }%
    }%
}
\makeatletter
\newcommand*{\ensquared}[1]{\relax\ifmmode\mathpalette\@ensquared@math{#1}\else\@ensquared{#1}\fi}
\newcommand*{\@ensquared@math}[2]{\@ensquared{$\m@th#1#2$}}
\newcommand*{\@ensquared}[1]{%
\tikz[baseline,anchor=base]{\node[draw,outer sep=0pt,inner sep=0.6mm,minimum width=4mm] {#1};}} 
\makeatother

\definecolor{ourgreen}{rgb}{0.00,0.49,0.19}
\definecolor{ourred}{rgb}{0.77,0.03,0.11}
\definecolor{ourorange}{rgb}{0.89,0.45,0.13}
\definecolor{ourgrey}{rgb}{0.60,0.60,0.60}
\def\yes{\textcolor{ourgreen}{\large\checkmark}}
\def\maybe{\textcolor{ourorange}{\Large$\circ$}} 
\def\no{\textcolor{ourred}{\Large\texttimes}}
\def\unknown{\textcolor{ourgrey}{\encircled[1mm]{?}}}

%% file: include/introduction.tex
\section{Introduction}
\label{sec:intro}

Many applications rely on commodity \glspl{tee} to protect data \emph{in use}. 
Technologies such as Intel \gls{tdx} and AMD \gls{sev}-\gls{snp} adopt a cloud-centric threat model: they assume an attacker can control every layer of host software and defend against passive physical attacks through full-memory encryption, but explicitly exclude active hardware tampering such as bus interposition ~\cite{eisoldt2025sokcloudyviewtrust}. 
Under that assumption, the \gls{tee}'s remote attestation certifies the CPU model, microcode, and measured launch state, but it carries no evidence of \emph{where} the processor is installed. 
A determined operator who does have physical custody can therefore migrate or, by default, operate an otherwise certified workload onto hardware in an uncontrolled environment, leaving the verifier with no cryptographic way to detect this change. 

Commercial \glspl{tee} are designed to protect against a malicious operating system, yet they implicitly depend on the platform owner to provide truthful information about the underlying hardware environment. This tension reveals a subtle but important gap in prevailing \gls{tee} threat models that has already had real world implications. The most recent attacks against \glspl{tee} leveraged physical access to the host machine to extract Intel \gls{sgx} attestation keys~\cite{wiretap,batteringramsp26,tee-fail} and also affect Intel \gls{tdx}. While such attacks remain out of scope under the threat models assumed by commercial \gls{tee} vendors, a remote third-party verifier that interacts with a \gls{cvm} only through a network-facing API has no independent means to validate where the \gls{tee} is physically deployed, nor to detect whether the attested execution is being relayed or proxied from a different location using standard attestation mechanisms.

Nevertheless, many attacks do not require physical access and instead exploit the hypervisor to observe or infer sensitive execution patterns~\cite{heraclesCPA2025,tdxdownCCS24}. 
This threat is especially challenging in cloud deployments, where tenants have no visibility on the hypervisor, an assumption that is commonly considered acceptable under \glspl{tee} threat models but is gradually contradicted by recent works~\cite{snpeek2025,schlüter2024hecklerbreakingconfidentialvms}.

This gap between what attestations prove and what users need to verify is particularly relevant when the party whose data is at stake is not the one operating the \gls{tee}. 
A genomics firm, an AI inference provider, or an online bank may process confidential data inside \glspl{cvm} hosted on a reputable cloud platform, but its customers have no cryptographic means to confirm this.
The service operator knows where its workloads run, but the end user does not. 
In \gls{defi}, where deployments increasingly rely on \glspl{tee} to protect high-value assets~\cite{8806762,Rabimba_2021}, the problem is worse since participants are mutually untrusted and often lack a stable, verifiable identity. 
In both cases, standard \gls{tee} attestations do not bridge this gap: the end user has no way to verify which infrastructure operator hosts the service, and therefore no basis on which to evaluate the physical trust guarantees of the execution environment. This missing notion of environment provenance falls outside today's \gls{tee} threat model. In the remainder of this paper, we treat environment provenance as a first-class security objective and introduce \gls{dcea} to address it.

We argue that \gls{tee} attestations lack explicit infrastructure binding and thus propose extending them to offer \gls{dcea}, a novel approach that cryptographically associates a confidential workload not only with a vetted software and hardware state, but also with a known infrastructure environment. 
This is achieved by combining two roots of trust - one provided by the \gls{tee} manufacturer and the second by the infrastructure owner, e.g., in the form of a \gls{tpm}. 
We prototype the manufacturer root of trust on Intel \gls{tdx}, as the only platform currently supporting both \gls{cvm} and bare-metal modes in \gls{gcp}. 
For the second root of trust, we establish a trust chain from the physical \gls{tpm} (or cloud \gls{vtpm}) to the \gls{cvm}, utilizing the provider's root of trust and its respective certificate chain. 
Here, we assume that the \gls{tpm}'s \gls{ekc} is issued by the cloud provider. \gls{dcea} generalizes to any \gls{tee} offering comparable attestation primitives.

Essentially, \gls{dcea} enshrines the data center as a disinterested, mutually trusted third party, on the economically rational assumption that mounting a chassis-level attack would damage the provider's reputation and future revenue far more than any benefit it might gain. We therefore develop a solution for two settings: the common managed-\gls{cvm} and the more demanding bare-metal deployment. The bare-metal scenario also provides additional security guarantees regarding the hypervisor, as it verifies its integrity. Both scenarios, along with their attack surfaces, are detailed in \Cref{sec:design}.

Our work has the following key contributions (KCs):
\begin{enumerate}[label={\bfseries KC\arabic*}, leftmargin=0.87cm]
\item We formalize the notion of environment provenance for \glspl{cvm}, identifying a critical blind spot in current \gls{tee} threat models regarding physical platform residency and infrastructure binding (\Cref{sec:analysis});

\item We design and implement the \gls{dcea} protocol, which utilizes a novel in-guest binding mechanism to cryptographically link \gls{tee} reports with platform-level \gls{tpm} quotes, anchoring a workload to a specific physical chassis (\Cref{sec:design});

\item Using the \gls{guc} framework and the AGATE model, we formally prove that \gls{dcea} provides location-binding guarantees, illustrating its resilience against advanced proxy attacks (\Cref{sec:formal-guarantees});

\item We evaluate a concrete \gls{dcea} instantiation on bare-metal Intel \gls{tdx} instances, demonstrating that it mitigates a broad range of software-level adversarial actions (A1--A6), including measurement forgery and channel tampering with minimal performance overhead (\Cref{sec:protocol-impl}).
\end{enumerate}

%% file: include/background.tex
\section{Background}
\label{sec:background}
This section introduces relevant background information detailing the two roots of trust on which \gls{dcea} relies: \gls{tee} attestation flows and \gls{tpm}/\gls{vtpm} quote mechanisms. For \glspl{tee}, we focus especially on Intel \gls{tdx}, which will be the target of our prototype implementation, and touch on AMD \gls{sev}-\gls{snp}.

\begin{figure}
    \centering
    \includegraphics[width=.8\columnwidth]{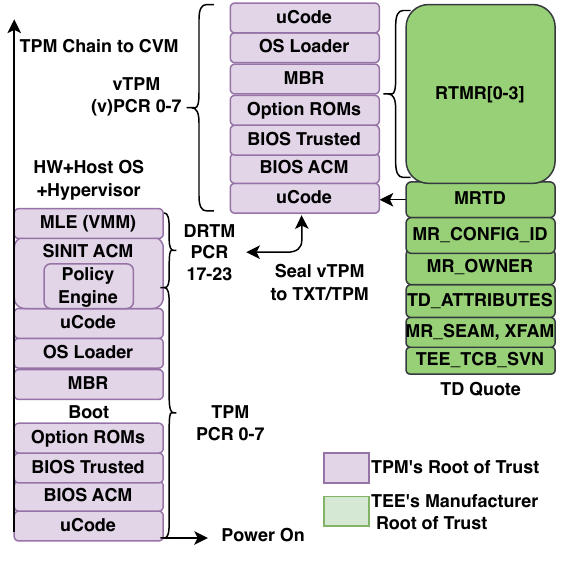}
    \caption{Overview of the host \gls{pcr} stack using \gls{tpm} and \gls{drtm} solution e.g., Intel \gls{txt}. Intel \gls{txt} extends \glspl{pcr} values to the hypervisor. Later, we bind the \gls{vtpm} to the host OS \glspl{pcr} and seal the \gls{ak} to the \glspl{pcr}. The \gls{td} attestation values overlap with \gls{vtpm} \gls{pcr} values~\cite{Futral2013,10.1145/3652597}. The stack does not cover all fields.}
    \label{fig:pcr-stack-complete}
\end{figure}

\subsection{Trusted Platform Module \& Measured Launch}

The \gls{tpm} enhances platform security by providing hardware-rooted measurement, attestation, and key management primitives. During the boot process, the \gls{crtm}, typically implemented in firmware, initiates a sequence of integrity measurements over early boot components. These measurements are extended into the \gls{tpm}'s \glspl{pcr}, forming a cryptographic chain of trust over the platform launch state~\cite{ShepherdTrustcom}.

A central element of the \gls{tpm}'s attestation capability is the \gls{ek}, a unique asymmetric key pair provisioned by the manufacturer~\cite{Stefan2006vTPM}. The public portion of the \gls{ek}, often distributed with an \gls{ekc}, serves as a hardware root of trust. In our setting, we assume that the \gls{ekc} is issued by an authority capable of certifying both the authenticity and physical security of the platform. To preserve privacy and limit key exposure, \glspl{tpm} derive \glspl{ak} from the \gls{ek}. These \glspl{ak} are used to sign attestation data, such as selected \gls{pcr} values, without directly revealing the \gls{ek}, while the accompanying \gls{ekc} enables verification of the trust chain back to the issuing authority.
In virtualized environments, the \gls{vtpm} extends \gls{tpm} functionality to individual virtual machines, providing each \gls{vm} with an isolated logical instance of \gls{tpm} services~\cite{Stefan2006vTPM}. The \gls{vtpm} maintains a per-\gls{vm} \gls{pcr} state reflecting guest boot measurements, as illustrated in \Cref{fig:pcr-stack-complete}; we therefore sometimes refer to this structure as the (v)\gls{pcr} stack. \glspl{vtpm} support common \gls{vm} lifecycle operations such as suspend, resume, and migration by securely preserving and transferring this state. However, the security of a \gls{vtpm} is fundamentally coupled to the trustworthiness of the underlying hypervisor, and a compromised host can undermine its isolation guarantees~\cite{Stefan2006vTPM}. Recent designs, such as Intel \gls{vtpm}~\cite{intel_vtpm_td} and the COCONUT SVSM \gls{vtpm}~\cite{coconut-svsmvtpm2026Jan}, mitigate this risk by executing the \gls{vtpm} within the \gls{tcb} of the \gls{cvm}, but do not by themselves bind the \gls{cvm} to a specific host software stack.

\gls{drtm} technologies complement the \gls{tpm} by extending the platform’s chain of trust beyond static firmware measurements and into the operating system or hypervisor launch. As a concrete example, Intel \gls{txt}~\cite{intel_txt_whitepaper} performs a hardware-enforced measured launch in which critical launch components, including firmware, boot loaders, and the initial hypervisor state, are measured and extended into additional \gls{pcr} registers. This enables detection of unauthorized modifications to the launch configuration and allows cryptographic keys to be sealed to a specific, verified platform state. In such deployments, the \gls{tpm} provides the persistent hardware root of trust, while the \gls{drtm} mechanism ensures that the resulting \gls{pcr} values accurately reflect the host software stack on which higher-level abstractions, including confidential virtual machines, are built.

\subsection{Trusted Execution Environments}

\begin{figure}
    \centering
    \begin{subfigure}{0.49\columnwidth}
    \centering
    \includegraphics[width=.5\columnwidth]{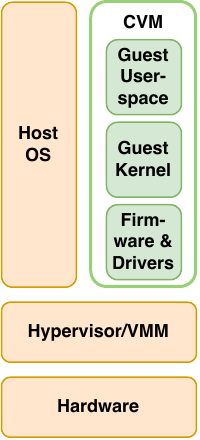}
    \caption{Bare/CVM Flow}
    \label{fig:bare-flow}
    \end{subfigure}
    \begin{subfigure}{.49\columnwidth}
    \centering
    \includegraphics[width=.73\columnwidth]{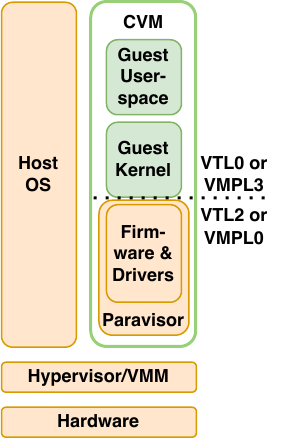}
    \caption{Paravirtualization Flow}
    \label{fig:para-flow}
    \end{subfigure}
    \caption{Simplified TEE attestation flow for various deployments. The provider controls the grey dotted boxes. For Bare/CVM flow, all relevant components are running in the trusted area (green) within the \gls{cvm}. In the case of Paravisor, certain components of Paravisor and its firmware are not trusted (orange) as they might not always be open-source.}
    \label{fig:attest-flows}
\end{figure}

\gls{vm}-based \glspl{tee}, such as Intel \gls{tdx}~\cite{inteltdx3:online,9448036} or AMD \gls{sev}-\gls{snp}~\cite{ménétrey2022exploratory,AMDSEVGit,li2022sokteeassistedconfidentialsmart}, enhance \gls{vm} security through encrypting and isolating guest \glspl{vm} from the hypervisor and supporting nested virtualization. 
Users gain confidence in a given \glspl{tee} enclave via the request of a remote attestation. 
For \glspl{cvm},~\Cref{fig:attest-flows} presents two attestation flows varying between bare-metal/native virtualization (\ref{fig:bare-flow}) and with an additional paravirtualized layer (\ref{fig:para-flow}), and how UUID is available to \gls{cvm}.
In the bare-metal setup (\ref{fig:bare-flow}), the \gls{cvm} runs directly on the hypervisor, e.g., QEMU. 
Attestation in this scenario involves verifying the firmware, operating system, and \gls{tee} itself. 
On the other hand, in the paravirtualized environment (\Cref{fig:para-flow}), the \gls{cvm} additionally relies on a paravisor, e.g., OpenHCL~\cite{microsoft2025Feb} or COCONUT~\cite{coconut-svsm2025Feb}.
The paravisor enables live migration of the \gls{cvm} and provides an additional layer of virtual drivers between the guest OS and underlying \gls{vmm}.

The paravisor implements an access mode as a \gls{vtl} for Intel \gls{tdx} and as a \gls{vmpl} for AMD \gls{sev}-\gls{snp}. 
Of note, \gls{vmpl}0 is the highest privilege level, and \gls{vtl}0 is the lowest, hinting at other implementation approaches.
The attestation report should include verification of the same components as regular \gls{cvm} flow deployment and the paravirtualization logic.
The paravisor's logic typically includes a lightweight hypervisor and additional drivers that map to the underlying hypervisor, as shown in~\Cref{fig:para-flow}.
This requires the paravisor's components to be open source to enable reproducible builds and obtain the checksum to compare with the value in the attestation fields. 
This is, however, not always the case, as was the case of Microsoft Azure's paravisor before OpenHCL~\cite{microsoftConfidentialAzure}. 
In addition, the measurements of the \gls{vtpm} used for the paravisor and the \gls{cvm} cover different components, which cannot be matched.
Therefore, the binding between the hardware provider and \gls{cvm} cannot be ensured.
Even when using the paravisor approach, the quote contains the \gls{ppid} constructed during Intel's initial platform provisioning phase. 
The \gls{ppid} is derived during the platform's registration with Intel's verification service. 
More importantly, \gls{ppid} is unique to the CPU and every \gls{td} has access to it.

\subsubsection{Intel TDX Attestation}

In the context of Intel \gls{tdx} \glspl{cvm} are referred to as \glspl{td}. 
A \gls{td}'s remote attestation enables external parties to verify the integrity and authenticity of a \gls{td} and provide information about the code running inside of it. 
The attestation report, or "quote," generated during this process contains several fields that provide insights into the \gls{td}'s configuration and state.

For \gls{dcea}, we need to understand the trust assumptions of the attestation process and have an overview of host-controlled components. 
We are especially interested in the \gls{td} quote fields and the information they provide. 
The \gls{tdx} attestation process relies on a root of trust established through interaction with Intel, which provides two Intel \gls{sgx} enclaves and a certificate chain verifying the authenticity of the platform. 
The flow begins with the \gls{tdqe}, a special-purpose \gls{sgx} enclave responsible for generating a \gls{td}-specific asymmetric key pair.
The \gls{tdqe} creates an \gls{sgx} report containing its identity and a hash of the \gls{ak} public key.
This report is verified locally by the \gls{pce}, another \gls{sgx} enclave also signed by Intel, using symmetric keys established via \gls{sgx}'s local attestation mechanism on the host OS. 
The \gls{pce} then derives the \gls{pck} key from platform-specific \texttt{EGETKEY} instructions, which reflect the host's \gls{tcb} level and package identity.
The \gls{pce} signs the report from the \gls{tdqe}, thus generating a certificate that vouches for the authenticity of the \gls{td}'s attestation key. 
This process is rooted in Intel’s \gls{pcs}, which issues X.509 \glspl{pckcert} binding the public part of the \gls{pck} key to the host platform.

The resulting quote, signed by the \gls{tdqe} using its \gls{ak}, includes measurements of the \gls{td} and can be verified against the \gls{pckcert} chain provided by Intel. 
This mechanism ensures that the quote was produced by a genuine \gls{td} running on a legitimate Intel \gls{tdx} platform. 
However, running a \gls{td} and operating on a legitimate platform does not indicate understanding of where the host operates. 
Therefore, we introduce \gls{dcea}, which combines Intel's checks with extensions to the environment in which the platform operates to strengthen trust assumptions.  
The \gls{td}'s memory and execution context are isolated from the hypervisor and host OS via the \gls{seam}, which mediates sensitive operations and enforces \gls{tdx}-specific protections. 
Memory encryption is handled by \gls{mktme}, which derives symmetric keys per-\gls{td} from hardware-resident secrets and configuration specific to \gls{tdx}. 
Overall, the attestation root of trust in Intel \gls{tdx} builds on \gls{sgx} enclaves, \gls{pck}, and \gls{pcs}, forming a hardware-rooted, verifiable trust chain for remote attestation.

The attestation report includes fields such as \texttt{TEE\_TCB\_SVN}, which indicates the security version number of the \gls{tee}'s \gls{tcb}, and \texttt{MR\_SEAM}, providing the measurements of the \gls{seam} module.
For more details, there are also \texttt{SEAMATTRIBUTES}, detailing the configuration of the \gls{seam} module, and \texttt{TDATTRIBUTES} field representing the attributes of the \gls{td}.
These attributes mainly aim to verify the expected security settings.
Measurements like \texttt{MRTD} (\gls{mrtd}), \texttt{MROWNER}, and \texttt{MROWNERCONFIG} provide cryptographic hashes representing the \gls{td}'s initial state, configuration, owner identity, and owner-specific configuration. 
\glspl{rtmr} are captured in \texttt{RTMR[0]} through \texttt{RTMR[3]}, reflecting dynamic components loaded after \gls{td} initialization. 
Last, \texttt{report\_data} is a 64-byte field supplied by the \gls{td}, often used to include nonce or other verifier-specific data~\cite{intel_trust_authority_attestation_tokens, intel_tdx_whitepaper_2022}. 
\Cref{fig:pcr-stack-complete} introduces simplified \gls{td} attestation report as the \gls{td} is initialized.



\subsubsection{Binding vTPM and TD}
\label{sec:binding-tdx-tpm}
Combining details from both \gls{pcr} and \gls{td}, we show in \Cref{tab:cross-arch-mapping-mapping} the mappings of the crucial values enabling our proposed protocol to work. 
Notably, these mappings might not match out of the box, as the \gls{vtpm} might not be able to provide the same values as the \gls{tdx} quote.
This can be mitigated by running dedicated logic inside the \gls{td} that extracts the \texttt{tpm\_eventlog} and provides the same values as the \gls{pcr} quote.
Unlike Intel's \gls{tdx}, which provides the \gls{rtmr} natively, we need to combine multiple \gls{pcr} values to match the \gls{vtpm} with \gls{rtmr} reports.
Importantly, the \gls{pcr}[0] is not sufficiently unique to provide strong binding.

As a consequence of the mapping covered in \Cref{tab:cross-arch-mapping-mapping}, we use Intel \gls{tdx} for our prototypes and illustrated examples. As we can see, AMD \gls{sev}-\gls{snp} offers a similar measured-launch flow called \texttt{LAUNCH\_DIGEST}, but it currently lacks \gls{rtmr}-style or comparable runtime integrity measurement capabilities~\cite{amd-sev-snp-whitepaper}. Our protocol would also be applicable to AMD \glspl{tee} provided \gls{sev}-\gls{snp} offers extended attestation flow in the future. 
Similarly, paravisor solutions, e.g., OpenHCL or COCONUT-SVSM, do not overlap with the guest OS's runtime and platform configuration, which is crucial for ensuring the binding between the platform and \gls{cvm}.
ARM \gls{cca}, on the other hand, provides \gls{rim} and \gls{rem}, with the same extend-only semantics as \gls{mrtd} and \gls{rtmr} measurements, respectively~\cite{arm-rmm-spec,li2022armcca}. 
The Linux kernel's TSM runtime measurement register abstraction already treats \gls{tdx} \glspl{rtmr} and \gls{cca} \glspl{rem} as first-class citizens behind a unified interface. 
Porting \gls{dcea} to \gls{cca} would thus require no changes to the protocol logic. 
However, no commodity \gls{cca} hardware exists yet to the best of our knowledge.

\begin{table}[h]
  \centering
  \setlength{\tabcolsep}{4pt}
  \caption{High-level mapping between \gls{tdx}, AMD SEV-SNP, and ARM CCA
           registers and guest vTPM \glspl{pcr}.
           Intel mapping based on~\cite{10.1145/3652597}.}
  \label{tab:cross-arch-mapping-mapping}
 \begin{tabular}{l p{2.2cm} ccc}
    \toprule
    \textbf{PCRs} & \textbf{Covered}
      & \textbf{Intel}
      & \textbf{AMD}
      & \textbf{ARM} \\
     & \textbf{Parts} & \textbf{TDX}
      & \textbf{SEV-SNP}
      & \textbf{CCA} \\
    \midrule
    0      & Virtual firmware (image)
      & \texttt{MRTD}
      & \texttt{LD}\textsuperscript{*}
      & \texttt{RIM} \\
    1,\,7  & Virtual firmware data \& config
      & \texttt{RTMR[0]}
      & ---
      & \texttt{REM[0]} \\
    2--5   & OS kernel, initrd, boot params
      & \texttt{RTMR[1]}
      & ---
      & \texttt{REM[1]} \\
    8--15  & OS apps / user-space integrity
      & \texttt{RTMR[2]}
      & ---
      & \texttt{REM[2]} \\
    ---    & Reserved
      & \texttt{RTMR[3]}
      & ---
      & \texttt{REM[3]} \\
    \bottomrule
    \multicolumn{5}{l}{%
      \textsuperscript{*}\,\footnotesize
      LD - \texttt{LAUNCH\_DIGEST}, Launch-time only.}
  \end{tabular}
\end{table}

%% file: include/analysis.tex
\section{Confidential VMs Security in Practice }
\label{sec:analysis}
This section highlights the discrepancy between the threat model of current \glspl{tee} implementations and the information provided in the attestations. 
We then dive into a detailed threat model that highlights the verifier's role and provides information about the execution location of the \gls{tee}.

\subsection{Physical-Access Gap in TEE Attestations}
\label{sec:physical_access_gap}

Existing attestation reports produced by Intel \gls{tdx} and AMD \gls{sev}-\gls{snp} certify the state of an enclave running on a genuine chip recognized by the vendor, but reveal nothing about the location of the hosting machine. 
A malicious operator can therefore serve an enclave and produce a valid attestation while secretly hosting the workload on hardware under their physical control. 
This becomes especially prominent with attacks that allow the extraction of attestation keys~\cite{wiretap,batteringramsp26}.
Our goal is to close this "location-oblivious" gap by binding the \gls{tee}'s attestation to an additional \gls{tpm} quote that authenticates the host as part of a public cloud-provider platform. 
With this discrepancy between the \glspl{tee} threat model and the attestation flow, we can see that the security gap stems from the trust in the operator to provide, but not to leverage their physical access to the machine to exploit side channel attacks. 
Today's commercially available \gls{cvm} solutions, e.g., Intel \gls{tdx} and AMD \gls{sev}-\gls{snp}, already raise the bar against a malicious hypervisor by encrypting memory, measuring launch, and enabling remote attestation, thereby minimizing host-software attacks on the \gls{vm} \cite{inteltdx3:online,ménétrey2022exploratory,AMDSEVGit,li2022sokteeassistedconfidentialsmart}. 
Nevertheless, recent work shows that even a malicious hypervisor can be used to extract secrets from the \gls{vm} memory~\cite{heraclesCPA2025}. 

\glspl{tee}' threat model implicitly considers different classes of physical adversaries. Passive physical attacks, such as cold-boot and offline DRAM analysis, are countered by full-memory encryption (TME-MK in TDX \cite{10.1145/3652597}, SME in SEV \cite{amd-sev-snp-whitepaper}). However, active bus interposition, as demonstrated by TEE.fail \cite{tee-fail} and \cite{batteringramsp26} at costs under \$1,000, remains explicitly out of scope per both Intel \cite{intel_tee_fail_reaction} and AMD \cite{amd_reaction} security advisories. The gap we address is therefore not that TEEs ignore physical access entirely, but that their attestation reports provide no evidence that the hosting environment is one where the excluded class of active physical attacks is physically prevented.


Cloud providers are incentivized by their reputation not to misbehave and attempt physical attacks on their own data center. Our threat model, therefore, embraces the data center side of the cloud provider as an element of the \gls{tcb}, providing physical protection to the \gls{tee} host's chassis and issuing valid certificates to its \glspl{tpm}. 
From the perspective of the \gls{cvm} and verifier, the threat model is only strengthened, as in an ideal scenario, we gain additional confidence about the provider.
If the \gls{vtpm} binding to the \gls{cvm} is compromised, we fall back to the default \gls{tee} threat model. We further expand on the threat model in \Cref{sec:threat-model}.
While \gls{tdx} and \gls{sev}-\gls{snp} significantly raise the bar for host-based attacks, they inherently rely on the CPU vendor (Intel or AMD) for implementation correctness, firmware integrity, remote attestation, and provisioning of encryption keys.
Even though recent work has shown that the \gls{tee} threat model is not as strong as it is assumed to be for \gls{cvm} and missing some crucial \gls{tcb} information and their verifiability~\cite{eisoldt2025sokcloudyviewtrust}.

As commonly assumed in commodity \gls{tee} literature, we treat the CPU vendor's key infrastructure as a single root of trust, and do not attempt to mitigate microcode backdoors or similar hardware vendor threats. 
Such supply-chain risks, while real, are orthogonal to the location problem addressed here.
We assume the machine is not compromised before it reaches the trusted provider location, and sensitive keys are cleared out in case the hardware is no longer in the protected location. 
Similarly, we do not consider sophisticated networking attacks, such as DNS spoofing. 
In summary, commodity \glspl{tee}' attestations prove "what code is running" but not "where it is running". 
By matching a provider-signed \gls{tpm} quote with the standard \gls{tee} attestation report, our protocol enables the verifier to confirm that the confidential \gls{vm} resides on cloud hardware outside the attacker's physical reach. 
This eliminates the last practical avenue for adversaries with physical access to conduct side-channel attacks in this setting.

\subsection{Threat Model}
\label{sec:threat-model}

\input{include/threat_model}

%% file: include/threat_model.tex
The primary objective of this threat model is to facilitate the secure deployment and interaction with confidential workloads by third parties in cloud environments. 
In these settings, tenants possess limited operational visibility. They rely exclusively on the hardware root of trust and the isolation provided by the \gls{tee}. The system must demonstrate attestation correctness, even if the platform operator behaves adversarially.

\subsubsection{Overview of parties and scenarios}

This threat model assumes that the cloud provider's physical infrastructure is trustworthy. It also assumes the provider does not engage in attacks that require physical access to the trusted execution environment \gls{tee} host, as described in \Cref{sec:physical_access_gap}.



The cloud provider, comprising the Host Manager \encircled{M} and Hardware Provider \encircled{P}, is modeled as an entity with a potentially adversarial software stack. Its measured launch process and certificate issuance infrastructure are assumed to be verifiable. The following behaviors are considered: (i) the provider is trusted to issue valid \gls{ekc} for its \glspl{tpm} and to provision \gls{ak} that are correctly measured and sealed under expected \gls{pcr} values; (ii) the provider is \emph{not} trusted to maintain confidentiality for data processed or stored in software-visible components, such as the \gls{vtpm}, which are fully accessible to the host operating system and hypervisor; and (iii) integrity guarantees are considered meaningful only when they are cryptographically verifiable, for example, through measured boot flows, sealed keys, and hardware-generated attestations.

\begin{figure}[t]
    \centering
    \includegraphics[width=.9\columnwidth]{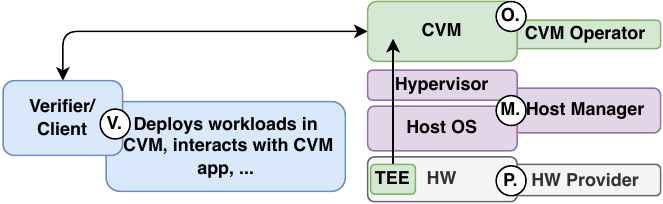}
    \caption{Overview of parties in the threat model. \textbf{Each party's color is used in subsequent figures as a line color}. Depending on the scenarios, some parties may overlap in their roles.}
    \label{fig:setting-threat-model}
\end{figure}

A critical requirement for the \gls{ekc} is issuance by an entity capable of certifying the infrastructure of a trusted provider, such as \gls{gcp} certifying an \gls{ek} within its own data center.
\Cref{fig:setting-threat-model} presents the overall process and identifies the individual parties involved. The following provides a detailed description of each party.
\begin{description}[leftmargin=1.5cm]
  \item[Verifier \encircled{V}] validates attestations and owns the workload's secrets. Also known as \textit{Client}. The Verifier wants to verify the integrity of the \gls{cvm} before deploying the workload and interacting with it.
  \item[CVM Operator \encircled{O}] the \textit{Tenant} who launches and manages the \gls{cvm} (may coincide with \encircled{V} in some cases).
  \item[The Host Manager \encircled{M}] is responsible for launching and managing the confidential host OS and hypervisor. In certain scenarios, the Host Manager role may overlap with or be fulfilled by the Hardware Provider \encircled{P}, particularly in deployment contexts where the environment is fully end-to-end provisioned and managed by the cloud provider.
  \item[HW Provider \encircled{P}] \textit{Cloud} infrastructure provider who controls physical servers, hypervisor, \gls{vtpm}/\gls{tpm}, and the networking layer.
\end{description}

Using these definitions, we can now refine our understanding of the cloud provider as a composite of the Hardware Provider \encircled{P} and the Host Manager \encircled{M}. We trust \encircled{P} for integrity and provenance (physical non-tamper and correct \gls{ekc} issuance), but do not trust any provider-controlled software (\encircled{M}) for confidentiality or channel honesty. Integrity holds only when cryptographically verifiable (measured launch, sealed keys, hardware-signed attestations).

The primary objective now is to provide assurance to the Verifier \encircled{V} that the workload executes within a trusted facility where hardware integrity is verified and remains uncompromised.

\subsubsection{Trust Assumptions}

\paragraph{Adversary Capabilities.}

We assume a strong software adversary that (i) controls the host OS, hypervisor, and virtualization stack (including any \gls{vtpm} instances); (ii) can intercept, delay, reorder, modify, drop, replay, or inject any messages exchanged between the tenant, verifier, and (v)\gls{tpm}; (iii) can provision arbitrary software on the host to masquerade as trusted infrastructure; and (iv) may collude with malicious insiders who possess host admin access.

\paragraph{Adversary Limitations.}

As per the commercial \glspl{tee}' threat model, we assume that CPU vendors (e.g., Intel, AMD) issue authentic microcode and root keys; the \gls{tpm} is genuine and has a valid \gls{ekc} signed by a public root; and the CPU enforces the correct isolation of \glspl{td} and implements \glspl{rtmr} as specified. Therefore, our design does not consider the hardware vendor as an active participant.

Assuming trust in the cloud provider's infrastructure, we rule out: physical tampering with CPUs, \glspl{tpm}, or server hardware, including chip extraction, side-channel sensors, or cold-boot attacks; compromise of the CPU or \gls{tpm} manufacturer's supply chain (e.g., issuance of fraudulent \glspl{ekc} or malicious microcode updates); attacks on hardware vulnerabilities unknown at deployment time (e.g., speculative execution exploits); and denial-of-service attacks, which may prevent availability but do not affect integrity. In summary, our threat model strengthens the standard \gls{tee} model by adding verifiable environment provenance: if the \gls{tpm} binding is compromised, security degrades gracefully to the baseline \gls{tee} guarantees.

With these assumptions, the adversary cannot forge hardware signatures, produce valid attestations inconsistent with the measured launch, or extract sealed keys from the \gls{tpm}. The following Sections \ref{sec:design} and \ref{sec:security-analysis} describe our proposed architecture and evaluate how it achieves the desired guarantees.

%% file: include/design.tex
\section{Architecture for DCEA}
\label{sec:design}

\begin{figure}[t]
    \centering
    \begin{subfigure}{0.48\columnwidth}
    \centering
    \includegraphics[width=.8\columnwidth]{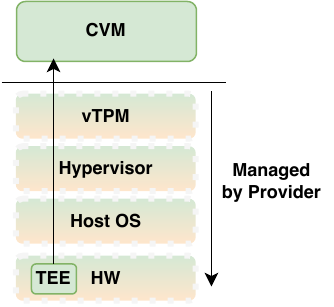}
    \caption{CVM with vTPM flow for cloud providers}
    \label{fig:cvm-vtpm-sol}
    \end{subfigure}
    \begin{subfigure}{.48\columnwidth}
    \centering
    \includegraphics[width=.78\columnwidth]{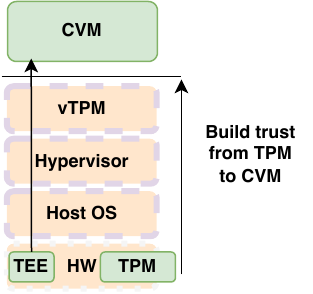}
    \caption{Bare-Metal with TPM building the trust to CVM}
    \label{fig:beremetal-tpm-sol}
    \end{subfigure}
    \caption{Overview of Scenario I (\textbf{S1}) and II (\textbf{S2}) introducing components within \gls{tcb} and outside of it, differentiated by the color code. Green corresponds to a trusted component and orange to an untrusted component. We assume for \textbf{S1} the components below \gls{cvm} as spatially trusted, as in many cases, the details and possible verification of the used source code are not possible. This differs for \textbf{S2}, where the components can be made deterministically open-source.}
    \label{fig:scenarios-tpm-cvm-flows}
\end{figure}

This section introduces how \gls{dcea} addresses the problem of binding a \gls{cvm} to its physical location under two deployment settings, illustrated in \Cref{fig:scenarios-tpm-cvm-flows}. An external verifier that interacts with the \gls{cvm} only through a remote API must be convinced that the \gls{cvm} executes in a specific physical location, typically a cloud data center. \textbf{Scenario I} (\textbf{S1}) considers a tenant \gls{cvm} running under a provider-managed hypervisor with a \gls{vtpm}, while \textbf{Scenario II} (\textbf{S2}) considers a single-tenant bare-metal deployment with access to a d\gls{tpm}.

\subsection{Scenario I: Confidential Virtual Machine}

In \textbf{Scenario I} (\textbf{S1}), we focus on a deployment in which the cloud provider manages both the host operating system and the \gls{vtpm} infrastructure, while the \gls{cvm} operator is protected by a \gls{tee}, as shown in \Cref{fig:cvm-vtpm-sol}. 
As per our threat model (\Cref{sec:threat-model}), the cloud provider is treated as a software adversary with full control over the host OS and \gls{vtpm}, but without the capability to alter the \gls{tdx} hardware root of trust or forge \gls{tpm} signatures as we see below.

We trust the provider to correctly provision the host and bind its \gls{ak} to the expected boot measurements. However, we make no confidentiality claims about material stored or handled by the \gls{vtpm}, and tenants must avoid placing secrets there.

\begin{figure}
    \centering
    \includegraphics[width=.9\columnwidth]{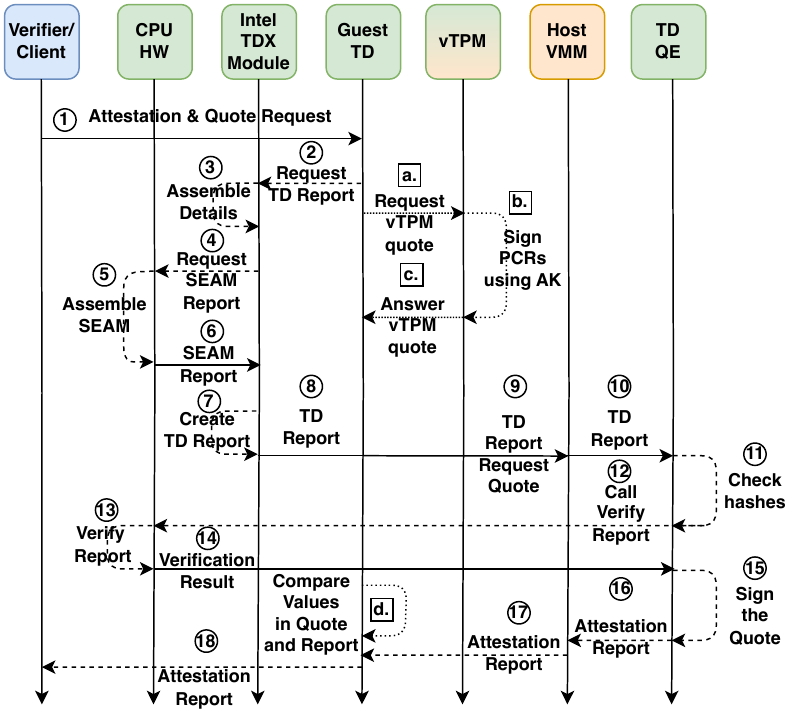}
    \caption{Sequence diagram of CVM TD attestation with a vTPM extension. Two parallel flows are shown: the vTPM flow (lettered square labels) and the TD attestation flow (circled numeric labels). Green boxes denote trusted components, while orange boxes denote malicious components.}
    \label{fig:cvm-vtpm-flow}
\end{figure}

The verifier interacting with the \gls{cvm} aims to bind the \gls{cvm} to a trusted infrastructure location. To achieve this, the \gls{vtpm} quote and the \gls{cvm} quote are compared. The \gls{cvm} can request a quote from the \gls{vtpm}, which includes measurements of the boot process recorded in \gls{pcr} registers. At the same time, the \gls{cvm} attestation report issued by the trusted execution environment includes runtime integrity measurements of the isolated execution environment. The process of obtaining the attestation from the isolated domain and the \gls{vtpm} quote is shown in \Cref{fig:cvm-vtpm-flow}. We further illustrate the attestation flow by using Intel \gls{tdx} as a concrete instantiation, showing how both the \gls{vtpm}'s \gls{pcr}-based measurements and the TDX's hardware-anchored \gls{rtmr} measurements are combined into the composite attestation evidence.

First, the client/verifier requests (\encircled{1}) the \gls{tee} attestation and \gls{vtpm} quote from the \gls{cvm}. 
There are two parallel flows - the \gls{td} attestation (\encircled{\#}) and the \gls{vtpm} quote (\ensquared{x}).
The \gls{td} requests a quote from the \gls{vtpm} (\ensquared{a}) and the attestation from the Intel \gls{tdx} module (\encircled{2}).
The \gls{vtpm} signs the current \gls{pcr} values using its \gls{ak} and returns the quote to the \gls{td} (\ensquared{b}-\ensquared{c}).
Concurrently, the Intel \gls{tdx} module collects \gls{seam} related platform measurements (\encircled{5}), which are then forwarded to construct a \gls{seam} report via the CPU HW layer itself, which then communicates with the \gls{tdx} module (\encircled{6}-\encircled{8}).
This report contains the \gls{td}'s configuration and runtime state and is passed to the host \gls{vmm} (\encircled{9}), which relays it to the Intel-signed \gls{td} \gls{qe} (\encircled{10}).
The \gls{td} quote enclave verifies the contents of the \gls{td} report by checking hashes and measurement values (\encircled{11}-\encircled{12}).
Once validated, the \gls{td} quote enclave generates and signs a final attestation report (\encircled{13}-\encircled{16}), which is returned to the \gls{td} (\encircled{17}).
The \gls{td} or verifier can then compare the \gls{vtpm} quote with the \gls{td} quote enclave-signed report to ensure that the \gls{vtpm} measurements and \gls{td} runtime state are consistent and bound to the correct platform (\encircled{18}).
Performing this check \emph{inside} the \gls{td} is especially attractive: only the boolean outcome leaves the enclave, so the verifier learns the binding’s validity without exposing sensitive metadata embedded in the \gls{vtpm} quote.

Although the \gls{vtpm} quote and the \gls{td} quote are produced by different roots of trust, they overlap on a common set of measurements.
\glspl{pcr} 0–7 in the \gls{vtpm} quote correspond to the \glspl{rtmr} in the \gls{td} quote.
Because \glspl{rtmr} are fixed in hardware once the \gls{td} launches, the cloud provider cannot alter them. 
Consequently, any forged or proxied \gls{vtpm} quote that reports \gls{pcr} values inconsistent with the \glspl{rtmr} will be detected immediately by the \gls{cvm} operator when the two artifacts are compared.

It is worth noting that the \gls{vtpm} does not enhance the integrity of the \gls{td} itself, as the \gls{td} quote is already rooted in Intel’s \gls{tdx} hardware and measured launch process. 
While the \gls{vtpm}'s \gls{ak} is certified and its measurements verifiable, we do not rely on the \gls{vtpm} for storing or handling confidential tenant material, as it is operated by the host and thus visible to the provider.
The general role of the \gls{vtpm} is to provide an additional binding layer: it enables secure provisioning of cryptographic keys, sealing of configuration, and verification of the host stack through sealed \glspl{ak} and certified \glspl{ek}.
However, we only strengthen the overall trust model by reinforcing the consistency between the internally provided \gls{cvm} \glspl{rtmr} and \gls{pcr} values provided by \gls{vtpm}.
The \gls{cvm} operator can verify that these two views align, building confidence in the platform’s state without needing to distrust the host OS. 
This model underscores the importance of integrating \gls{cvm} quotes with \gls{tpm}-based attestation when the host \gls{os} is trusted, but the cloud provider still wishes to enable verifiable binding. 



\subsection{Scenario II: Bare-Metal Deployment}

In \textbf{Scenario II} (\textbf{S2}), a tenant executes a \gls{cvm} on a bare-metal server that additionally exposes a \gls{dtpm} to the underlying host (\Cref{fig:beremetal-tpm-sol}). The Host Manager \encircled{M} installs its own host OS and hypervisor, thus making the software stack beneath the guest \emph{untrusted}, whereas the chassis remains in the provider's physically secured data center. 
None of the host-side software stack, including the OS, hypervisor, or any \gls{vtpm} process, is trusted. Any guarantee must be based solely on hardware.
Our goal in this setting is to provide an external verifier with the same level of assurance it enjoyed in \textbf{S1} about the \gls{cvm}'s physical location, while tolerating a fully adversarial host stack. 
The verifier must be able to confirm that the \gls{cvm}'s runtime state, the \gls{vtpm} instance, and the physical platform all belong to a single, untampered machine inside the cloud provider's data center.

As a concrete instantiation, we describe the protocol using Intel \gls{tdx} and Intel \gls{txt}, while noting that the design generalises beyond this specific platform. Other architectures that provide a CPU-rooted attestation report and \gls{tpm}-backed key sealing can realise the same construction without modification to the protocol logic. In this instantiation, the protocol anchors trust in Intel's measured-boot facility, \gls{txt}. During system startup, \gls{txt} extends measurements of the firmware, kernel, and the \gls{vtpm} binary into \gls{pcr} 17-18 of the platform \gls{tpm}. Because the \gls{tpm} authenticates the exact host software stack via these \glspl{pcr}, the \gls{vtpm}'s \gls{ak} can be sealed to the measured launch state. Finally, the \gls{td} hashes the \gls{ak}'s public key into its attestation report (for example, the \texttt{report\_data} or \texttt{MRCONFIGID} field). This binds guest evidence to the sealed \gls{vtpm}, forming a chain of trust that links the guest OS, the TPM-sealed attestation key, and the measured platform state.

From the verifier's perspective, the message flow is identical to \textbf{S1}: it still requests a \gls{cvm} report and a \gls{vtpm} quote. The underlying chain of evidence is, however, stronger. Hardware alone vouches for every layer, so even if the host OS is operated by a different party from the hardware owner, the verifier can detect any attempt to swap, replay, or forge components. 
The following sections analyze the adversarial attack surface of \gls{dcea}, systematically evaluating each attack vector summarized in \Cref{tab:attack-summary} and establishing the protocol's security guarantees through formal argument.

%% file: include/security_analysis.tex
\section{Security Analysis}
\label{sec:security-analysis}

\subsection{Security Goals}
\label{sec:security-goals}

Our design aims to provide the verifier with clear and enforceable guarantees against a powerful host-level adversary. At a high level, the verifier must be convinced that attestation evidence (\gls{tee} report and \gls{vtpm} quote) originates from genuine roots of trust, is cryptographically bound to a single physical platform, and cannot be substituted, replayed, or proxied from another machine. The evidence must remain fresh, consistent across \gls{tee} and \gls{tpm} measurements, and verifiable even when transmitted over host-controlled channels. We exclude threats such as physical tampering or supply chain compromise, as outlined in our threat model.

These goals apply in both deployment scenarios. In \textbf{S1}, they ensure provider-operated \glspl{vtpm} are properly provisioned and tightly linked to the \gls{cvm} launch state. For \textbf{S1}, the provider controls both the vTPM and the \gls{ekc} issuance, so the binding rests on the provider's operational integrity rather than on hardware sealing alone—a weaker but still economically rational guarantee, as discussed in \Cref{sec:threat-model}. In \textbf{S2}, they ensure \gls{dtpm} quotes and \gls{cvm} attestations both come from the same bare-metal chassis, even with an untrusted host stack. Together, these objectives form the basis for analyzing the concrete attacks next.

\begin{figure}[t]
    \centering
    \begin{subfigure}{\columnwidth}
    \centering
    \includegraphics[width=.7\columnwidth]{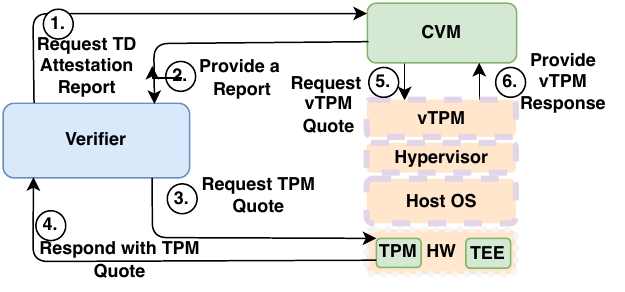}
    \caption{Naive Attestation Flow with honest host. In this case, the verifier can trust the \gls{tpm} quote and the \gls{td} quote.
It \encircled{1} requests the \gls{td} report and receives a report \encircled{2} which can be properly verified.
Next, the verifier \encircled{3} sends a request to the host \gls{os} to get the \gls{tpm} quote, for which the host sends a response \encircled{4} back. 
If using the \gls{vtpm} quote as in \encircled{5}, the verifier can provide a nonce to the \gls{cvm} \texttt{report\_data} attestation and the request of the \gls{vtpm} quote as \encircled{5}, which is then part of the \gls{vtpm} response.}
    \label{fig:baremetal-naive}
    \end{subfigure}
    \begin{subfigure}{\columnwidth}
    \centering
    \includegraphics[width=.8\columnwidth]{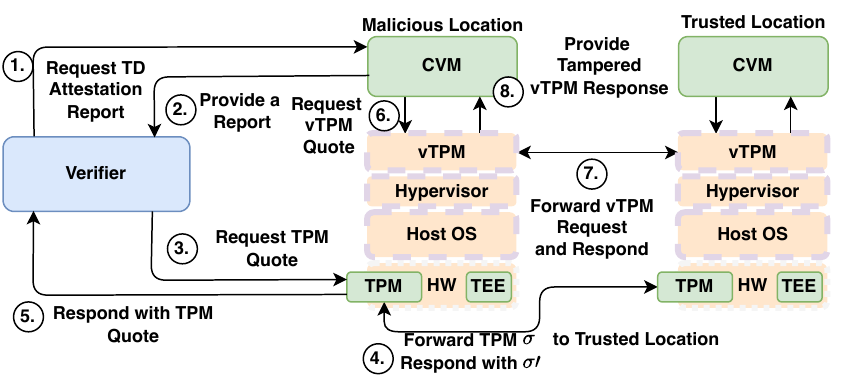}
    \caption{"Mix-and-Match" Proxy Attack. The malicious host can modify the \gls{tpm} quote to impersonate a trusted platform. Quote requests and responses are now relayed through the attacker's infrastructure to simulate a valid attestation chain.}
    \label{fig:beremetal-Frankenstein}
    \end{subfigure}
    \caption{Overview of naive and malicious attestation flows from the perspective of the verifier.}
    \label{fig:challenge-verifier-flow}
\end{figure}

To clarify how our protocol addresses each threat, we detail the naive attestation flow shown in \Cref{fig:baremetal-naive}. A verifier confirms host integrity by requesting a nonce-bound \gls{tee} report and a matching \gls{tpm} quote. Identical digests and shared nonces prove the artifacts originate from the same correctly booted platform. An adversary controlling the host OS (but lacking physical access) aims to deceive the verifier regarding the \gls{cvm}'s cloud location using software-based attacks. This threat is most acute in bare-metal deployments (\textbf{S2}), where the host stack is untrusted. We examine six attacks (A1-A6) that a malicious Host Manager \encircled{M.} may attempt.

\subsection{Possible Attack Strategies}
\label{sec:attacks}

In the naive attestation flow, the verifier confirms host integrity by matching nonces and digests from a \gls{tee} report and a \gls{tpm} quote. An adversary controlling the host (but lacking physical access) aims to deceive the verifier regarding the \gls{cvm}'s location. We categorize the primary vectors (A1–A6), establishing the proxy as the central threat that relies on subsequent primitives.

\paragraph{A1: Relay \& Proxy (Primary Threat).} The adversary initiates an attack by relaying quote requests to a separate, honest machine to simulate a valid trust chain.
In the sophisticated variant, the attacker launches the \gls{td} locally but proxies all \gls{tpm} interactions to a remote cloud host. 
This covert attack succeeds only if the adversary can synchronize nonces and reproduce matching \gls{pcr} digests and hashes across two different platforms.

\paragraph{A2: Quote Forgery.} The provider may attempt to forge or simulate the entire \gls{vtpm} quote or inject falsified \gls{pcr}/\gls{rtmr} values. 

\paragraph{A3: Measurement Inconsistency.} A naive proxy (A1) typically results in a detectable mismatch where the remote machine's \glspl{pcr} do not align with the local \gls{td}'s \glspl{rtmr}. The adversary's goal is to mask this inconsistency to avoid being detected. 

\paragraph{A4: Channel Interception.} To stage A1 or A2, the malicious \encircled{M.} must compromise connection between the \gls{cvm}, \gls{vtpm}, and \gls{dtpm}. 

\paragraph{A5: Identity Substitution.} The adversary may spoof platform identity by reusing or faking valid \gls{ek} certificates from legitimate hardware (e.g., issued by Intel or a cloud provider) to mislead the verifier. They may also substitute the expected \gls{ak} with a different key under their control to break the cryptographic link between the \gls{td} and the \gls{tpm}.

\paragraph{A6: Component Compromise.} The host may instantiate a modified or malicious \gls{vtpm} binary that signs attacker-controlled measurement values or leaks cryptographic material. Such privilege-level compromises are often necessary to facilitate proxying or forgery while maintaining a facade of trustworthiness.

\begin{table*}[htb]
  \centering
    \caption{Summary of adversarial actions (A1--A6) with applicability across Scenario I (trusted host OS and vTPM operated by the cloud provider) and Scenario II (untrusted host requiring cryptographic attestation for the vTPM and host stack). \yes~means the attack is relevant, \no~means the attack is not relevant.}
  \begin{tabular}{c p{4.6cm} p{6.8cm} c c}
    \toprule
    \textbf{A\#} & \textbf{Category name} & \textbf{Representative actions} &
      \textbf{Scenario I} & \textbf{Scenario II} \\
    \midrule
    A1 & Relay/proxy ("mix-and-match") &
          Proxy quote to another machine; proxy; Synchronized replay of nonce/PCRs &
         \no & \yes \\[2pt]    
    A2 & Quote/measurement forgery &
         Forge or simulate vTPM quote; Inject falsified PCR/RTMR values &
         \yes & \yes \\[2pt]
    A3 & Measurement inconsistency &
         PCRs $\neq$ TD‑RTMRs; \texttt{report\_data} mismatch &
         \yes & \yes \\[2pt]
    A4 & Channel interception/tampering &
         MITM on TD to vTPM path; MITM on host to pTPM bus &
         \no & \yes \\[2pt]
    A5 & Identity \& key substitution &
         Spoof TPM‑EK certificate; Replace expected AK &
         \no & \yes \\[2pt]
    A6 & Privilege‑level component compromise &
         Run modified or malicious vTPM binary &
         \no & \yes \\
    \bottomrule
  \end{tabular}

  \label{tab:attack-summary}
\end{table*}

\paragraph{Summary}
While \textbf{Scenario I} (provider-managed) is vulnerable to specific subsets of these threats, \textbf{Scenario II} (bare metal) exposes the broadest attack surface, necessitating a protocol that explicitly defends against relay and substitution attacks to enable trustworthy bare-metal deployments. 

%% file: include/formal_proof.tex
\subsection{Formal Guarantees}
\label{sec:formal-guarantees}

To demonstrate that \gls{dcea} provides robust guarantees against relay and proxy attacks, we formalize the security of the \gls{dcea} protocol using the \gls{guc} framework. Specifically, we adopt the AGATE model \cite{agate2024}, which proposes a modular blueprint for representing different classes of TEE functionality within its $G_{att}^{mod}$ framework.

This approach allows us to model channel corruptions and the adversary $\mathcal{A}$'s control over the host OS while preserving universal composition guarantees. We explicitly model \gls{dcea} as a \textbf{Wrapper Protocol} $Prot_{DCEA}$ that bridges the gap between the standard TDX functionality $G_{TEE}$ and an ideal functionality $G_{CloudTEE}$ capable of attesting to its cloud location. The $\mathcal{G}_{CloudTEE}$ functionality is defined as the modular instantiation:
$$\mathcal{G}_{CloudTEE} := \mathcal{G}_{att}^{mod}\left[\lambda, \mathcal{R}_{Intel}, \mathbb{O}_{CloudTEE}, \mathbb{A}_{TEE}, \mathbb{S}_{TEE}\right]$$

Where $\mathbb{O}_{CloudTEE} = \mathbb{O}_{TEE} \cup \{GetLocationProof\}$.

Upon enclave installation, the $\mathcal{G}_{CloudTEE}$ functionality initializes a new shell $sh_{CloudTEE}$ that captures its behavior:

\begin{shell}[label=func:sh_cloudtdx, title={$sh_{CloudTEE}[\msf{prog}]$}]

\heading{Parameters.}
Program $\msf{prog}$.

\heading{Interfaces.}
$\mathbb{O} = \mathbb{O}^{std} \cup \{\msf{Quote}, \msf{GetLocationProof}\}$,
$\mathbb{A} = \emptyset$.

\heading{Extended Identity.}
$(sh_{CloudTEE}[\msf{prog}], (eid || pid, \msf{att} || idx))$.

\Init
\begin{itemizeless}
  \item Initialize empty set of virtual \glspl{iti}.
\end{itemizeless}

\OnInput \inmsg{\msf{INSTALL}} from $\mathcal{G}_{att}^{mod}$:
\begin{itemizeless}
  \item \If virtual ITI $(\msf{prog}, (eid, idx))$ does not exist:
        \begin{itemizeless}
          \item Create virtual ITI $(\msf{prog}, (eid, idx))$.
        \end{itemizeless}
  \item \If ideal functionality $(\mathcal{G}_{TEE}, (idx, \bot))$ does not exist:
        \begin{itemizeless}
          \item Create ideal functionality $(\mathcal{G}_{TEE}, (idx, \bot))$.
        \end{itemizeless}
  \item \textit{(The shell ensures the existence of the helper functionality it relies on.)}
\end{itemizeless}

\OnInput \inmsg{\msf{inp}} from $\mathcal{G}_{att}^{mod}$:
\begin{itemizeless}
  \item Execute program $(\msf{prog}, (eid, idx))$ step by step on input $\msf{inp}$.
  \item For each instruction $i$:
        \begin{itemizeless}
          \item \If $i \in \mathbb{O}^{std}$:
                \begin{itemizeless}
                  \item Allow standard execution of $i$.
                \end{itemizeless}

          \item \Else \If $i = \msf{GetLocationProof}(\msf{nonce})$:
                \begin{itemizeless}
                  \item \Send \inmsg{\msf{LOC\_PROOF}, \msf{nonce}}
                        to $\mathcal{G}_{TEE}$.
                  \item Receive response $\lambda$.
                  \item Write $\lambda$ to the subroutine output of
                        $(\msf{prog}, (eid, idx))$.
                \end{itemizeless}

          \item \Else \If $i = (\textbf{return } v)$:
                \begin{itemizeless}
                  \item \Output
                        $v$ with source
                        $(sh_{CloudTEE}[\msf{prog}], (eid || pid, \msf{att} || idx))$.
                \end{itemizeless}
        \end{itemizeless}
\end{itemizeless}

\end{shell}

Our primary security goal is to ensure \emph{Location Binding}. Informally, if a verifier accepts a composite attestation generated by \gls{dcea}, then the \gls{cvm} execution and the platform \gls{vtpm} quoting mechanism must originate from the same physical platform, subject to the hardware assumptions defined in~\Cref{sec:threat-model}.
This is captured in Theorem \ref{thm:cloudtdx} and the associated proof sketch.
The reader will find the full proof and set of functionalities in Appendix \ref{app:formal-model}.

\begin{theorem}
\label{thm:cloudtdx}
Let $\mathcal{G}_{TEE}$ denote the vanilla instance $\mathcal{G}_{att}^{mod}[\dots, sh_{TEE}]$ and
let $\mathcal{G}_{CloudTEE}$ denote the target instance $\mathcal{G}_{att}^{mod}[\dots, sh_{CloudTEE}]$.
Then the protocol $Prot_{DCEA}$, which installs the wrapper shell $W^{DCEA}$,
in the presence of $\mathcal{G}_{TEE}$ UC-emulates $\mathcal{G}_{CloudTEE}$ against any static adversary
$\mathcal{A}$ controlling the Host.
\end{theorem}

\paragraph{Proof Sketch.}
Our proof relies on the reduction properties of the AGATE framework and proceeds in four main steps.
\subparagraph{\textbf{Step 1: Reduction via AGATE composition.}} We invoke AGATE \textit{Theorem 6} (General Replacement of Global Setups), which allows us to demonstrate that our wrapper protocol $Prot_{DCEA}$ combined with the weaker setup $\mathcal{G}_{TEE}$ emulates the stronger $\mathcal{G}_{CloudTEE}$ setup.
\subparagraph{\textbf{Step 2: Simulator construction.}} We design a simulator that manages an internal simulation of the \gls{tpm} functionality $\mathcal{G}_{TPM}$ to enforce \gls{pcr} states and applies AGATE's signature transformation to make real-world signatures indistinguishable from ideal-world.
\subparagraph{\textbf{Step 3: Handling the proxy attack.}} The simulator enforces consistency between the execution shell and the \gls{vtpm}; if $\mathcal{A}$ attempts a proxy attack (relay quotes from a remote node), the simulator detects the mismatch or the failure to unseal keys due to \gls{pcr} divergence, mirroring the hardware sealing failure in the ideal world.
\subparagraph{\textbf{Step 4: Indistinguishability.}} Finally, we demonstrate that no environment can distinguish the real execution from the ideal simulation, relying on the unforgeability of hardware-backed signatures and the correctness of the wrapper's binding logic.

%% file: include/protocol_mitigations.tex
\section{Protocol Implementation and Attack Mitigations}
\label{sec:protocol-impl}

This section introduces a candidate implementation of the \gls{dcea} protocol, which complements the architecture described in \Cref{sec:design}. The \gls{dcea} design mitigates the attacks outlined in \Cref{sec:attacks} through a four-step process: i) establishing trustworthy platform roots and a measured launch; ii) provisioning and sealing attestation keys to the measured state; iii) binding in-guest evidence across the \gls{tee} and the \gls{vtpm}; iv) and defining the verifier workflow for validating composite evidence over host-mediated channels. \Cref{fig:implementation-flow} depicts the stack components involved in each step. As a reminder, the threat model in \Cref{sec:threat-model} assumes a software-only adversary who controls the host OS, hypervisor, and all communication channels but lacks physical access. CPU/\gls{tpm} roots (and their supply chains) are trusted.

\subsection{Four-step DCEA Protocol Implementation}

We prototyped \gls{dcea} on \gls{gcp}, a public cloud offering that supports measured launch anchored in \glspl{pcr} 17 and 18 via Intel \gls{txt} and Intel \gls{tdx}, to demonstrate feasibility. This configuration enables the protocol to realize the bare-metal \textbf{S2}, which is strictly stronger than \textbf{S1} and shares the same attestation flow. We introduce $T_{DCEA_{Total}}$ in~\Cref{eq:model}, modelling the total time overhead of \gls{dcea} protocol in practice, with results shown in~\Cref{fig:time_boxplot}. Overall, we see that \gls{dcea} is practical, especially for bare metal deployments, introducing low overhead in addition to used \gls{tee} attestation flow, as outlined in Appendix~\ref{app:dcea-performance}. Additional implementation, requirements, and platform availability details are provided in Appendix \ref{app:implementation}.

\begin{figure}[ht!]
    \centering
    \includegraphics[width=.7\columnwidth]{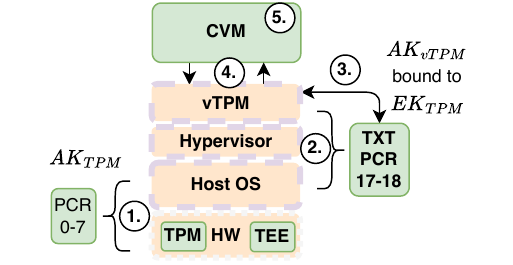}
    \caption{End-to-end DCEA flow. \encircled{1} shows what is covered by \gls{tpm} quote with \gls{txt} extending the PCR all the way to hypervisor \encircled{2}, while relying on the same $AK_{TPM}$. \encircled{3} continues with generation of $AK_{vTPM}$ against \glspl{pcr} values bound to $EK_{TPM}$. The \gls{cvm} interacts with the \gls{vtpm} and uses the $AK_{vTPM}$ in the process, \encircled{4}, and handles verification \encircled{5}.} 
    \label{fig:implementation-flow}
\end{figure}

Following the~\Cref{fig:implementation-flow}, we unfold the individual steps:

\textbf{Steps \encircled{1} \& \encircled{2} - Measured Launch Anchor:} A secure boot resets the \gls{tpm} and measures early boot and host OS into \glspl{pcr} 0–7. These \glspl{pcr} form a hardware-backed summary of the host stack (provider-managed in \textbf{S1}, bare-metal in \textbf{S2}) and serve as the root for policy and verification. The Intel \gls{txt} measures the hypervisor, \gls{vtpm}, and an attestation agent within the host's \gls{tcb}, and extends \glspl{pcr} 17-18. The \gls{dtpm}'s quote uses \gls{ak} certified by \gls{dtpm}'s \gls{ek}, which then bears the data center's root of trust certificate chain. As part of the flow, clear policies for the host OS and hypervisor expectations must be defined.

\textbf{Step \encircled{3} - AK Provisioning and Sealing:} The \gls{vtpm} creates an \gls{ak} whose usage is policy-bound to \glspl{pcr} 17–18. The \gls{ak} can sign quotes only when the platform matches the approved measured state, with provenance tied to the platform \gls{ekc}, preventing replay or key substitution. 

\textbf{Step \encircled{4} - In-Guest Binding:} The \gls{td} embeds the hash of the v\gls{tpm}’s public \gls{ak} in its attestation report and verifies quotes under that \gls{ak}. The v\gls{tpm} \gls{ak} is created with a \gls{pcr}-bound usage policy, ensuring it can only be exercised when the host is in the expected \gls{txt}-measured state. In addition, the attestation agent within the measured \gls{tcb} obtains a binding statement from the physical \gls{tpm}, whereby the physical \gls{ak} signs a digest of the v\gls{tpm} \gls{ak} public key. Together, the policy-bound v\gls{tpm} \gls{ak} and the physical \gls{tpm} binding signature transitively anchor the \gls{td}’s evidence to the measured host stack and the platform’s \gls{ekc}, preventing mix-and-match compositions or key substitution.

\textbf{Step \encircled{5} - Composite Verification:} The verifier checks freshness, validates the \gls{td} report and \gls{vtpm} quote, confirms the embedded $hash(AK_{pub})$, and ensures \glspl{pcr} 17–18 align with \gls{td} measurements. \gls{ekc} and \gls{td} chains establish hardware binding, detecting forgery, replay, and proxying. The extraction of the \gls{dtpm}'s material happens over a secure channel established with the attestation agent running in the \gls{txt}-measured \gls{tcb}.

\subsection{Attack Mitigations}
\label{sec:methodology:attack-mitigations}

Our security analysis centers on the proxy attack (\textbf{A1}), where a malicious host proxies \gls{tpm} interactions to a remote machine. This acts as a superset of the other adversarial capabilities. As such, successfully mitigating A1 inherently neutralizes the building blocks that enabled it in the first place (A2--A6).

\paragraph{Preventing Preparation (A2, A4, A6).}
To stage A1, the adversary must intercept channels (\textbf{A4}) or modify host components (\textbf{A6}). The \gls{dcea} protocol renders interception futile by enforcing end-to-end integrity through hardware-bound signatures. Furthermore, any modification to the \gls{vtpm} binary or host OS (\textbf{A6}) alters the launch measurements in \gls{pcr}. Because \gls{ak} is sealed to this state, the modified stack loses access to signing keys, rendering forgery (\textbf{A2}) impossible.

\paragraph{Role of PCR Values in DCEA.}
We stress that \gls{pcr} and \gls{rtmr} values serve as \emph{software-stack consistency checks}, not as unique machine identifiers. 
Two platforms running identical firmware, kernel, and hypervisor versions will produce identical \gls{pcr}\,0--7 digests by design.
Machine-level uniqueness in DCEA is instead established by the \gls{ekc} chain, which binds attestation evidence to a specific physical \gls{tpm} certified by the cloud
provider and the \gls{tdx}'s attestation containing a unique hardware ID.
The \gls{pcr}/\gls{rtmr} cross-check ensures that the \gls{cvm} and the vTPM share the \emph{same} measured state on that certified chassis, closing the mix-and-match vector (A1) where an attacker might combine a legitimate \gls{ekc} from one machine with \gls{cvm} execution on another.
In short, the \gls{ekc} answers \emph{which machine}, while the \gls{pcr}/\gls{rtmr} match answers \emph{on the same machine}.

\paragraph{Enforcing Binding and Identity (A3, A5).}
If the adversary relies on substitution (\textbf{A5}) rather than valid hardware, the In-Guest Binding (\textbf{Step~3}) exposes the deception. The \gls{td} embeds the hash of the local \gls{vtpm}'s public key ($hash(AK_{pub})$) in its report. If the host relays a quote from a remote machine with a different identity, this hash will mismatch. Finally, the verifier cross-checks the \gls{td}'s \glspl{rtmr} against the quoted \gls{pcr}s; any discrepancy triggers a measurement inconsistency (\textbf{A3}), immediately exposing the attack.

The protocol defeats the proxy relay (A1) through strict hardware binding. Since the \gls{ak} is sealed to the physical platform's launch state (\glspl{pcr} 17--18), it cannot be used by a proxy on unmatched hardware.
\input{include/new_table_proof_attacks_mapping}
\Cref{tab:mitigation-proof-mapping} maps the specific attacks to their concrete system mitigations and the corresponding mechanisms in the formal UC proof to summarize the last two Sections \ref{sec:security-analysis}-\ref{sec:protocol-impl}.

%% file: include/new_table_proof_attacks_mapping.tex
\begin{table*}[t]
  \centering
  \caption{Consolidated view of DCEA security: Mapping the primary proxy threat (A1) and its supporting primitives (A2--A6) to system mitigations and formal security model mechanisms.}
  \label{tab:mitigation-proof-mapping}
  \renewcommand{\arraystretch}{1.2}
  \begin{tabular}{p{0.20\linewidth} p{0.36\linewidth} p{0.36\linewidth}}
    \toprule
    \textbf{Attack Scope (ID)} & \textbf{System-Level Mitigation (Concrete)} & \textbf{Formal Model Mechanism (Abstract)} \\
    \midrule
    
    \textbf{Relay \& Proxy} \newline (A1: Proxy) & 
    \textbf{Freshness \& Channel:} Nonces are injected into both TD Report and TPM Quote. & 
    \textbf{Challenge-Response:} The ideal $\mathcal{G}_{CloudTDX}$ requires nonce $n_V$ for location proofs, which the simulator $\mathcal{S}$ enforces via the signature transformation. \\
    \midrule

    \textbf{Forgery \& Compromise} \newline (A2, A4, A6) & 
    \textbf{TXT Sealing:} The AK is sealed to PCR 17--18. The hardware enforces that the key is unusable if the host stack (A6) or vTPM binary differs from the measured launch state. & 
    \textbf{$\mathcal{G}_{TPM}$ Access Control:} The global functionality enforces specific \texttt{Policy} checks over the simulated \texttt{PCRs}. If the adversary state differs, \texttt{TPM2\_Quote} returns $\bot$. \\
    \midrule
    
    \textbf{Inconsistency \& Substitution} \newline (A3, A5) & 
    \textbf{Crypto Binding \& Identity:} The TD embeds $hash(AK_{pub})$ in its report. The verifier validates the EK certificate chain against the manufacturer or provider CA. & 
    \textbf{$W^{DCEA}$ Binding Logic:} The wrapper explicitly computes binding $B$ and aborts if it mismatches the vTPM output. $\mathcal{G}_{TPM}$'s registry $\mathcal{R}_{Root}$ is initialized only with valid EKs. \\
    
    \bottomrule
  \end{tabular}
\end{table*}

%% file: include/discussion.tex
\section{Discussion}
\label{sec:discussion}

\subsection{Comparing Scenarios: Challenges \& Limitations}

The protocol, threat model, and analysis of potential attacks and mitigations identify significant trade-offs and considerations when deploying \gls{dcea} systems in multitenant cloud environments. 
Both scenarios, \textbf{S1} and \textbf{S2}, leverage similar attestation flows but differ largely in flexibility, trust assumptions, and control granularity.

Functionally, both scenarios employ a similar attestation flow, leveraging \gls{cvm} quotes, cryptographic binding to \gls{ak}, and remote verifiability of \gls{pcr} values. \textbf{S2}, when available on the underlying cloud, is the preferred deployment model due to its greater control, flexibility, and auditability. \textbf{S2} requires that additional components of the stack, such as the hypervisor and BIOS, should be made open-source to enable third parties to verify the stack details. 
With a fully managed bare-metal setup, tenants can enforce per-\gls{cvm} unique identifiers, customize \gls{pcr} measurement policies, and benefit from boot-time integrity guarantees all the way to the hypervisor. 
Although this approach increases complexity, such as requiring extra logic to be deployed in the \gls{cvm} to evaluate quote freshness or binding, it offers stronger assurance and is more resilient to replay or proxy-based attacks, as well as most side-channel attacks that can be executed from the hypervisor. 

A practical limitation lies in the dependency on \gls{pcr} measurement flows. 
These may differ across platforms or virtualization stacks, especially in \textbf{S1}, where the tenant has no control over the \gls{vtpm} instantiation or the way (v)\gls{pcr} are appended. 
For \textbf{S2}, such configurations are at least inspectable by the Host Manager \encircled{M} using a specific \gls{vtpm} implementation.
The \gls{pcr} value interpretation, especially for higher-indexed registers, may not always reflect a uniform boot flow unless explicitly standardized across providers~\cite{constellation_attestation}.
This discrepancy may create challenges when comparing the \gls{pcr} values with runtime measurements in the \gls{cvm}, making reproducibility harder in such cases.
When measurements do not match, a mechanism must be implemented to reconstruct the sequence of events that led to a given \gls{pcr} population. 

This could be resolved if \gls{tee} vendors provided a \gls{vtpm} as a part of the \gls{tcb}. In the specific case of Intel \gls{tdx}, the \gls{vtpm} would live in the TD module, for instance. Such a solution would unify the deployments across different providers and offer greater security guarantees than hypervisor-based \glspl{vtpm}.

\subsection{Offering Unique Identities in Multitenancy}

Ideally, each \gls{cvm} should be bound to a uniquely provisioned \gls{ak} in multitenant environments. This requirement is especially critical in environments where tenants cannot control or verify host configurations.
In practice, it is possible to reuse \gls{vtpm} components and their respective \gls{ak}. A single \gls{vtpm} certificate chain may be shared across multiple \glspl{cvm} on the same physical host. While technically feasible, this weakens the uniqueness guarantees of attestation and may complicate the detection of identity cloning or proxy attacks. 

To enhance accountability and enforce the uniqueness of \glspl{ak}, a global or domain-scoped registry of \gls{ak} public keys could be established. 
This registry could be implemented using a \gls{ct}~\cite{google_certificate_transparency}, or registered on a public blockchain. 
Each newly provisioned \gls{ak} would be registered, along with metadata such as the issuing entity, timestamp, and, optionally, a \gls{cvm} identifier or platform attestation. 
Verifiers could then consult this registry to detect duplicate \glspl{ak}, ensuring that the \gls{cvm} is not impersonating nor cloning already provisioned identities. 
Additionally, the \gls{cvm} itself could verify during boot or session setup whether its \gls{ak} is unique and consistent with expected infrastructure bindings. 
Such a registry would increase transparency and allow audits to detect violations of identity guarantees across tenants or providers. In single-tenancy deployments with one \gls{cvm} per hardware node, verifying the public registry would ensure that only one \gls{ak} and its corresponding \gls{ek} are present, confirming the expected behavior.

\subsection{Privacy Considerations}

The \gls{vtpm}’s certificate chain often encodes deployment details such as cloud region and availability zone \cite{salrashid123_gcp_vtpm_ek_ak}.
Forwarding the entire \gls{vtpm} quote to the verifier would therefore expose information the tenant may wish to keep private. 
In the flow of \Cref{fig:cvm-vtpm-flow}, the \gls{cvm} already holds both artifacts, so the obvious alternative is to perform the consistency check inside the \gls{cvm} and release only a single-bit result.
A possible solution to retain public verifiability without leaking the chains is to generate a \gls{zkp} to convince the verifier that first, the \gls{vtpm} quote and \gls{tee} attestation report are consistent, and second, the certificate chain of \gls{vtpm} is included in a pre-defined set of approved cloud providers.

Concretely, cloud vendors would maintain a list of root certificates from which the \gls{vtpm} certificate chain is derived. The prover could then show set inclusion via a Merkle tree membership proof \cite{cryptoeprint:2019/1255}.
The same approach applies to \gls{ppid}-based solutions \cite{intel_poe_2025}, where the cloud provider lists the \glspl{ppid} running in their infrastructure, and the verifier can query it.
We do not impose any specific requirements on the proof system. The resulting circuit should fit comfortably within \gls{cvm} memory. The main challenge this method would likely face is convincing cloud providers to publish the input material required as part of the \gls{zkp} extension.

\subsection{Related Work}

A particular research avenue focuses on location-binding attestations for \gls{dcea}, aligning with our threat model that aims to ensure computations occur within specific, trusted environments~\cite{cryptoeprint:2021/697, cryptoeprint:2023/999, rezabek2025narrowinggapteesthreat}. 
Existing solutions typically rely on latency-based verification or external geolocation signals such as GPS or cellular networks. 
However, these approaches either suffer from noise-induced inaccuracies on the network path or require additional hardware that may be exposed to the \gls{cvm} and lack the required precision~\cite{cryptoeprint:2021/697}. 
In contrast, our approach builds on already available infrastructure, enabling extensibility within existing deployment strategies.

Our threat model introduces the \emph{proxy} attack, where a malicious host attempts to convince a verifier that a \gls{cvm} is running in a trusted environment when it is not. The proxy attack is similar to the concatenation attack introduced in~\cite{shang2024ccxtrustconfidentialcomputingplatform}, but it expands the adversarial capabilities to include attackers who own and coordinate multiple physical platforms. 
In this setting, individual attestation components—while valid in isolation—may be combined in a misleading manner to falsely assert trusted execution.
Several existing systems aim to harden individual components of the attestation pipeline. 
For instance, \texttt{vtpm-td}~\cite{intel_vtpm_td} mitigates attacks where a malicious hypervisor proxies or tampers with \gls{vtpm} interactions by running a dedicated \gls{vtpm} inside its own \gls{td}. 
Similarly, AMD SEV-SNP-based approaches, such as SVSM vTPM, place the \gls{vtpm} inside a higher-privilege in-guest security monitor, protecting the \gls{vtpm} state from host interference~\cite{coconut-svsmvtpm2026Jan}. 
These designs strengthen the confidentiality and integrity of \gls{vtpm}'s state, but remain focused on \emph{intra-VM trust}: they ensure that measurements reported by a \gls{cvm} are authentic, yet do not establish that the reported execution is co-located with a specific physical platform or owned infrastructure.

In contrast, our \gls{dcea} approach addresses \emph{inter-component binding} by cryptographically linking the \gls{cvm}'s launch and runtime integrity measurements to the host platform’s measured state. This binding directly counters proxy-style mix-and-match attacks and extends beyond protecting individual attestation primitives to ensuring their correct composition.
Other work has examined related proxy and replay vulnerabilities, such as the Time-Of-Check to Time-Of-Use (TOCTOU) problem in remote attestation~\cite{nunes2021toctouproblemremoteattestation}, where stale measurements can be reused after malicious state changes. 
While our solution does not implement continuous attestation, it significantly narrows the TOCTOU window by coupling the \gls{cvm} launch to the host’s measured state. 
Any post-launch changes in firmware or hypervisor configuration alter these measurements and are detectable upon re-attestation. Experiments in \Cref{app:dcea-performance} show that \gls{tpm} overheads are low, making periodic or continuous re-attestation feasible. Complementary mechanisms such as TGX~\cite{cispa2886}, which combines \gls{tpm} and Intel \gls{sgx} for secure key derivation, could further strengthen these guarantees.

From a platform ownership perspective, Intel's \gls{poe} \cite{intel_poe_2025} binds a platform identity to an organizational owner via signed endorsement tokens. Each platform exposes an immutable Processor Registration ID (PRID) and a resettable \gls{ppid}, embedded in attestation quotes and cross-referenced against provider-signed manifests. \gls{poe} and \gls{dcea} are complementary: POE answers which entity owns this hardware, whereas \gls{dcea} answers is the host software stack in a verified state.
Under \gls{poe} alone, a malicious provider (or a compromised insider) could run a modified hypervisor on valid, endorsed hardware to mount side-channel attacks like TDXdown, TDXploit, Heckler \cite{10.5555/3766078.3766141,tdxdownCCS24,schlüter2024hecklerbreakingconfidentialvms, snpeek2025}.
Because the roles are fulfilled by distinct entities with misaligned incentives, the Host-to-CVM interface becomes a critical attack surface. 
A combined deployment could use \gls{poe} for fleet-level identity and \gls{dcea} for per-CVM hypervisor integrity, providing defense in depth against both proxy attacks and host-stack compromise.

Prior work has highlighted concerns about the disclosure of sensitive information in \gls{ct} logs~\cite{9917530,Eskandarianprivacy}. For \gls{dcea}, we envision using \glspl{zkp} to allow a \gls{cvm} to prove membership in a set of trusted hardware identities without revealing the exact provider. Privacy-preserving mechanisms such as private set membership proofs or anonymous credential schemes could be used to mitigate traceability while preserving the strong binding guarantees provided by \gls{dcea}.
Finally, the \gls{dcea} concept is not limited to cloud deployments. A trusted certifying authority could similarly vouch for on-premise hardware by combining hardware-rooted attestation with tamper-evident enclosures and intrusion-detection records, analogous to Apple’s hardware-integrity program~\cite{apple_pcc_hardware_integrity}. This shows that \gls{dcea} represents a general architectural principle for execution-location assurance, rather than a cloud-specific mechanism.

%% file: include/conclusion.tex
\section{Conclusion}

In this paper, we addressed a long-standing gap in \glspl{cvm}' threat model: standard \gls{cvm} attestations prove \emph{what} code is running but not \emph{where}, leaving external clients exposed to hosts that run confidential workloads on hardware they control. 
For privacy-sensitive workloads, this gap has direct consequences: the physical environment where private data is decrypted and processed remains an unverified assumption.

We therefore adopted a threat model in which the entire host software stack may be adversarial. 
The external client, or verifier, is an independent party that needs hardware-rooted evidence of both code identity and physical location.
\gls{dcea} contributes three protocol-level building blocks to this problem that apply to a broad class of \gls{tee} and platform-attestation designs.
First, it layers \gls{tpm} quotes extended by \gls{drtm} measurements, and sealed \gls{vtpm} identities, to create a verifiable chain from chassis to guest, even when the host OS and hypervisor are malicious. 
Second, it introduces measurement-register cross-checks that defeat advanced relay attacks, including the newly described mix-and-match proxy attack. 
We formally prove these guarantees in the UC framework, showing that \gls{dcea} securely emulates an ideal execution-location oracle even in the presence of a malicious host software stack.
Third, we validate a reference implementation and reproducible measurement policies that run unchanged on today's \gls{tdx} hardware with Intel \gls{txt} and \glspl{dtpm}, demonstrating practical overheads. Together, these mechanisms defeat all six software-only attack vectors identified in our security analysis. 
The proxy attack subsumes the others and is the most demanding to counter. Mitigation follows from the hardware-enforced binding of the attestation key to the platform's measured launch state.
Even though we focus on \gls{tdx}, ARM \gls{cca} offers all required features, but no hardware is available to the public.
Unlike ownership or inventory-based endorsements such as Intel \gls{poe}, \gls{dcea} enforces runtime co-location between guest execution and platform state, and remains sound even when the hypervisor is adversarial.


The \gls{cvm} threat model is not one-size-fits-all: while many enterprise workloads assume that providers will not mount physical attacks, permissionless and adversarial settings are the norm in decentralized systems. 
These new settings demand cryptographic evidence of platform provenance. 
\gls{dcea} supplies this "Proof of Cloud" by binding guest attestation to provider-anchored measured boot, allowing remote verifiers to confirm execution on trusted data center hardware rather than on-premises or untrusted hosts. 
By making execution location a verifiable security property rather than an implicit assumption, \gls{dcea} enables confidential computing deployments that require verifiable platform provenance, from regulated enterprise workloads to adversarial, multi-tenant systems such as \gls{defi}.


%% file: include/implementation-appendix_arxiv.tex
\section{Implementation Details}
\label{app:implementation}
This appendix provides an overview of the implementation of \gls{dcea} for both \textbf{S1} and \textbf{S2} and lists fitting components for the implementation.

\subsection{Implementation Steps}

Following the architecture of \gls{dcea} outlined in \Cref{sec:design}, we observe the crucial role of several software or hardware components to mitigate possible threats. 
Therefore, we have to first investigate which platforms provide suitable support for these tools for both \textbf{S1} and \textbf{S2}. 
\Cref{tab:cloud_tpm_simple} summarizes a subset of known platforms and an overview of their Intel \gls{tdx} offerings and support for \gls{tpm}/\gls{vtpm}. 
At the time of writing, only \gls{gcp} offers solutions for both scenarios, which leads us to choose it.
Nevertheless, for \textbf{S1} we observe more offerings, as it requires only \gls{cvm} deployment of Intel \gls{tdx} and \gls{vtpm} support.
On the other hand, for \textbf{S2}, we need the bare metal offering and access to HW \gls{tpm}. 
Of note, if there is no offering of Intel \gls{tdx}, we give \no~to HW \gls{tpm}, too.  
We do not provide a dedicated column for Intel \gls{txt}, as is inherently available on Intel \gls{tdx} CPUs.
Building a secure attestation chain involves hardware \gls{tpm}, Intel \gls{txt}, \gls{vtpm}, and Intel \gls{tdx} to mitigate attacks outlined \Cref{tab:mitigation-proof-mapping}.
The main assumption for the adoption of \gls{dcea} is the extension of PKI solutions by cloud providers, which supports the issuance of \gls{ekc}.
The overall flow is introduced in \Cref{fig:implementation-flow}.

\begin{table}[ht]
\centering
\label{tab:cloud_tpm_simple}
\begin{tabular}{ l c c c c }
\toprule
 & \multicolumn{2}{c|}{\textbf{S1}}  & \multicolumn{2}{c}{\textbf{S2}} \\
\textbf{Platform} & \textbf{CVM} & \textbf{vTPM}  & \textbf{Bare metal} & \textbf{HW TPM} \\
\hline
AWS      &  \no & \yes & \no & \no \\
GCP      & \yes          & \yes    & \yes & \yes \\
Azure    & \yes & \yes    & \no    & \no \\
IBM Cloud &  \yes     & \maybe  &      \no        & \no \\
OVH      & \yes   & \no & \yes & \maybe \\
\bottomrule
\end{tabular}
\caption{Support Overview for Intel TDX, CVM/Bare‑Metal, vTPM, and HW TPM. \yes~full support, \no~no support for TDX, \maybe~special offerings}
\end{table}

\subsection{Envisaged Deployments}
As introduced, clients can interact with \glspl{cvm} or rely on cloud-managed bare-metal offerings, with access control over the host OS stack. 
Focusing on Intel \gls{tdx}, the \glspl{cvm} are becoming more widely present. 

\subsubsection{CVMs}
Major cloud providers vary in their support for Intel \gls{tdx}. 
\gls{gcp} and Microsoft Azure support Intel \gls{tdx} in both \gls{cvm} (e.g., GCP C3 instances~\cite{intel-gcp-confidential} and Azure DC/ECesv5-series~\cite{azure-confidential-preview}) and  bare-metal trust guarantees for \gls{gcp}.
Azure supports limited bare-metal deployments, not for Intel \gls{tdx}~\cite{azure-baremetal-overview}.
\gls{aws} offers AMD \gls{sev}-\gls{snp} based \gls{cvm}, but does not currently support Intel \gls{tdx} in either virtualized or bare-metal offerings. 
Even though \gls{aws} offers \gls{aws} Nitro, we do not cover it in this paper, as it is only offered by \gls{aws} by default.
\gls{gcp} provides an access to Intel \gls{tdx} and also \gls{vtpm} quotes through Google Cloud Attestation by running \gls{vtpm} client inside the \gls{cvm}~\cite{google-confidential-attestation}. 
The stack is based on QEMU with a \gls{vtpm} service, while attestation integrates both \gls{tpm} and \gls{tdx} attestation libraries such as \texttt{go‑tpm}~\cite{google-confidential-space, google-go-tpm}.
Azure also relies on Intel \gls{tdx}, and each \gls{cvm} includes its own isolated \gls{vtpm} \cite{azure-vtpm,azure-vtpm-leverage}. 
Same as for \gls{gcp}, the \gls{vtpm} exposes hardware-like \gls{tpm} interfaces. 
Attestation is managed through Azure Attestation and optional Intel Trust Authority integration~\cite{azure-vtpm}.
As introduced in \Cref{fig:para-flow}, Azure follows the paravisor approach.
This does not hinder the applicability of \gls{dcea}, as long as the v\gls{tpm} measurements can be bound to \gls{td}'s quote. 

For bare-metal deployments, Azure offers only AMD \gls{sev}-\gls{snp} \cite{azure-bare-metal}. 
\gls{gcp} is one of the few providers that offers bare-metal Intel \gls{tdx} servers~\cite{google-tdx-bare-metal}.
Therefore, we mainly focus on \gls{gcp} as it provides both \gls{cvm} and bare-metal Intel \gls{tdx} offerings.

\subsubsection{Host Stack Considerations}
In both \gls{gcp} and Azure \gls{cvm}, the host stack typically builds on established virtualization components (e.g., QEMU or hypervisor services) paired with a \gls{vtpm} service exposed via \texttt{vsock} or \texttt{virtio}. 
The \gls{vtpm} and hypervisor launching the \gls{td} are managed by the cloud provider and assumed to be measured and attested.
However, the \gls{cvm}'s operator has limited visibility of the underlying host manager's stack. 
On bare-metal systems, host managers have more control over the stack and gets OS configured for Intel \gls{txt} measurements and secure boot, paired with \gls{vtpm} integration via hypervisor of choice.

\subsubsection{Multi-Tenancy Solutions}

In multi-tenant deployments, for both a public-cloud setting or on a bare-metal cloud deployment, the requirement is a one-to-one binding between each workload’s trust boundary and the operator. 
For a \gls{cvm}, the identity is the \gls{vtpm}’s \gls{ak}.
In case of a bare-metal tenant, it is the physical \gls{tpm}’s \gls{ek}/\gls{ak} pair. 
The presence of a unique \gls{ak} per \gls{vtpm} instance ensures that any attestation token or quote can be traced to a single \gls{cvm}, irrespective of how many \glspl{td} or classical \glspl{vm} run on the same host. 
Without such uniqueness, cryptographic evidence loses its provenance: two \glspl{cvm} could present the same public key and therefore be indistinguishable to an external relying party, in case the workload is the same.

In the provisioning flow, the operator generates a fresh \gls{ak} for every \gls{vtpm} instance during \gls{cvm} instantiation.
The \gls{ak} can then be embedded in the \gls{td}’s \texttt{report\_data} (or encoded in a provider-signed \gls{vtpm} certificate), any remote verifier can bind the \gls{td}’s quote to that exact \gls{vtpm}. 
A theoretical hazard arises when the host software is allowed to launch \glspl{td} from arbitrary images without inspecting their embedded \gls{vtpm} state and respective \gls{pcr} values.
If the host manager \encircled{M} accidentally clones a snapshot that already contains an \gls{ak}, the new \gls{cvm} inherits the same key material.
Both tenants would then share a single attestation identity even though their runtime stacks may diverge, undermining auditability and potentially enabling cross-\gls{vm} replay of signed quotes. 
Hence, the platform’s control plane must provide a post-launch sanity check that rejects any instance whose \gls{ak} collides with an existing entry in the provider’s key registry. 
Overall, if the provider creates two identical \glspl{cvm} with the same \gls{ak}, it can be recognized.
This, in general, means that the security of higher-level protocols rests on the assumption that no two distinct \glspl{td} ever share the same \gls{ak}.

\section{Performance Evaluation of DCEA}
\label{app:dcea-performance}

The overall impact on the user interacting with~\gls{dcea} can be expressed using~\Cref{eq:model}. The $T_{Total_{DCEA}}$ corresponds to the total time, which includes time for retrieving the \gls{tpm} quote ($T_{Quote_{vTPM}}$), time to retrieve \gls{tee} attestation ($T_{Attestation_{TEE}}$), and their comparison ($T_{Match}$). We cover the evaluation of the contribution of individual components to $T_{Total_{DCEA}}$ in \Cref{fig:time_boxplot}. Additionally, time would be spent during boot time for both \gls{cvm} and the host OS when secure boot is enabled, but since this is a one-time event, we focus mainly on the runtime overhead.

\begin{equation}
    T_{Total_{DCEA}} = T_{Quote_{vTPM}} + T_{Attestation{TEE}} + T_{Match}
    \label{eq:model}
\end{equation}

\subsection{TPM Performance}
For our performance study of $T_{Quote_{vTPM}}$, we focus on evaluations of \gls{tpm} and \gls{vtpm} using both \gls{cvm} and bare-metal access within \gls{gcp}, hosted by OVH in the Canada region (based on public IP). 
We use a classical KVM-based \gls{kvm} instance and a single-tenant HW host. 
In each environment, we execute 500 consecutive operations against the platform’s root of trust. 
We operate in two areas: signing and quote generation. 
For both setups, we prepared the required key material, e.g., \gls{ak}. 
We interact with (v)\glspl{tpm} using the \texttt{tpm2-tools}.
Firstly, signing, for which we generate a plain text of \SI{64}{\byte}, which is sent to the \gls{tpm} before signing. 
When we issue quote requests, they read \gls{pcr} 0–7 before signing. 
The signature is over a hash value of individual \gls{pcr} values. 
We focus on SHA384, which matches the default hash function of Intel \gls{tdx}, and respective \gls{td}'s quote.
The end-to-end time is measured before the function is called and after it returns the response from \gls{tpm}.

\begin{figure}
 \centering
 \begin{subfigure}[b]{\columnwidth}
     \centering
     \includegraphics[width=\textwidth]{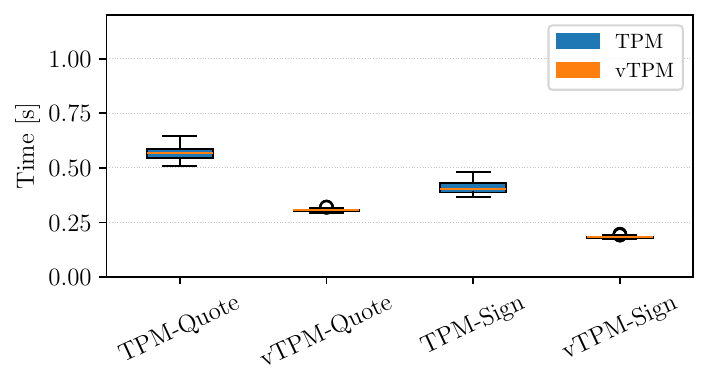}
     \caption{Execution time of the \gls{vtpm} and \gls{tpm} for the different configurations.}
     \label{fig:time_boxplot_tpm}
 \end{subfigure}
 \hfill
 \begin{subfigure}[b]{\columnwidth}
     \centering
     \includegraphics[width=\columnwidth]{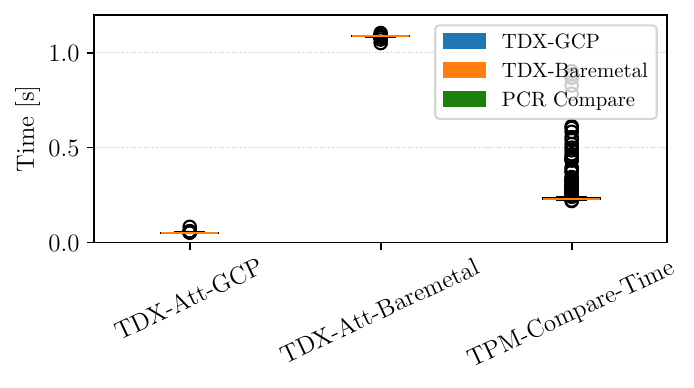}
     \caption{Execution time of the \gls{tdx} attestation in \gls{gcp} and local bare metal deployment and sample workload of \gls{pcr} compare time.}
     \label{fig:time_boxplot_tdx}
 \end{subfigure}
 \hfill
    \caption{Execution time of vTPM, TPM, and TDX attestation flow. Comparison of GCP and bare-metal deployments across different platforms. }
    \label{fig:time_boxplot}
\end{figure}


\Cref{fig:time_boxplot_tpm} shows the resulting distributions (Tukey box-plots). 
We compare both operations (Quote and Sign) in the same figure. 
On bare metal, when interacting with hardware \gls{tpm}, the quote centers around \SI{0.55}{\second}.
\gls{vtpm} is faster with average of \SI{0.3}{\second} per signature. 
The same pattern holds for signature generation, where signing is roughly \SI{150}{\milli\second} faster for both \gls{tpm} and \gls{vtpm}.
Overall, hardware \gls{tpm} takes longer, as it must serialize \gls{pcr} hash computations and authorization. 
We also observe more fluctuations, as \glspl{tpm} in general implement protections against timing attacks when following the requirements of the Trusted Computing Group~\cite{TCG_PTP_V1_06_2025}. 
We also observe larger outliers as a result. 

The data confirm two findings. 
First, hardware \gls{tpm} remains an order-of-magnitude slower than software \glspl{vtpm}.
Nevertheless, for our part, this is not an issue, as we do not expect regular calls; the \gls{cvm} operator should create the quote once for each client interaction. 
It is possible to cache the quote if no challenge is requested.
Second, we observe a clear pattern difference between \gls{tpm} and \gls{vtpm}, with \gls{vtpm} exhibiting much more stable performance. 

\subsection{TEE Performance \& Comparison}

Continuing with the evaluation of $T_{Attestation{TEE}}$ and $T_{Match}$, we consider two deployments of Intel \gls{tdx}. The first is a conventional \gls{cvm} hosted on \gls{gcp}, and the second is a bare-metal deployment using QEMU as the hypervisor and the Canonical \gls{tdx} flow~\cite{canonical-tdx}. To evaluate $T_{\text{Match}}$, we adopt a reconstruction logic for \gls{tpm} event logs, which is comparable to the process of binding \gls{rtmr} values to \glspl{pcr}. All experiments were conducted over 500 iterations.


\Cref{fig:time_boxplot_tdx} presents the evaluation results. For the \gls{gcp} deployment, the execution time per attestation is approximately \SI{0.05}{\second}. In contrast, the local bare-metal deployment requires around \SI{1.1}{\second}. This major difference is likely attributable to variations in host configuration, including potential caching or other platform-specific optimizations.

Finally, the \gls{tpm} comparison time includes extracting the event log and comparing it against the \gls{pcr} values stored in the \gls{tpm}. These steps are comparable to those required for mapping \gls{rtmr} values to \glspl{pcr}. While the runtime exhibits some fluctuations, the mean comparison time is approximately \SI{0.2}{\second}. Revisiting \Cref{eq:model}, a naive aggregation of the individual components results in $T_{{Total}_{DCEA}}$ of roughly \SI{0.6}{\second}. Depending on the deployment, either the \gls{tdx} attestation or the \gls{vtpm} quote is the dominant contributor to overall latency. Notably, \gls{tdx} attestation and \gls{vtpm} quote generation can be executed in parallel. Overall, these results demonstrate the feasibility of \gls{dcea}, with modest runtime overhead that can be further reduced through optimizations, such as caching selected \gls{vtpm} details.

%% file: include/appendix_dcea_proof.tex
\section{Formal Modelling: Modular Global Functionalities}
\label{app:formal-model}

We adopt the \gls{guc} framework to model the security of the \gls{dcea} protocol. 

\subsection{Modelling Rationale: AGATE Instantiation}

Following \cite{agate2024}, we model a generic \gls{tee} supporting \glspl{cvm}, such as Intel TDX or AMD SEV, as an instantiation of the Augmented Global Attested Trusted Execution functionality (AGATE) $G_{att}^{mod}$. We then refine this model to capture \glspl{tee} that additionally support location attestation, and contrast it with a baseline functionality representing standard \glspl{tee} that lack this capability. This formulation leverages the AGATE framework to express these extended behaviors while preserving the \gls{uc} guarantees of $G_{att}^{mod}$.

We subsequently model \gls{dcea} as a \textbf{Wrapper Protocol} ($Prot_{DCEA}$) that bridges the gap between the behavior of \gls{tee} and that of a hypothetical ideal \gls{tee} that could prove it runs in the cloud.

\subsection{The Global TEE Functionality ($\mathcal{G}_{TEE}$)}

The Global \gls{tee} Functionality is defined as the modular instantiation:
$$\mathcal{G}_{TEE} := \mathcal{G}_{att}^{mod}\left[\lambda, \mathcal{R}_{TEE}, \mathbb{O}_{TEE}, \mathbb{A}_{TEE}, \mathbb{S}_{TEE}\right]$$

where the parameters are defined as follows:

\begin{enumerate}
\item $\mathcal{R}_{TEE}$: The registry of valid platform signing keys ($PK_{CPU}$), representing the set of genuine hardware recognized by the manufacturer.
\item $\mathbb{O}_{TEE}$: The set of feature oracles available to the Guest \gls{tee}, defined as $\{ \texttt{ExtendRuntimeMeasurement}, \texttt{GetReport} \}$.
\item $\mathbb{A}_{TEE}$: The standard set of attacks possible against a vanilla \gls{tee} instance (side channels, etc.). We do not introduce additional adversarial capabilities, reflecting modern \glspl{tee}' standard threat model, where guarantees are provided against logical adversaries.
\item $\mathbb{S}_{TEE}$: The attestation generation function, defined to output a signature over the tuple $(\texttt{TEE\_{conf}} || \texttt{LaunchMeasurement} \\ || \texttt{RuntimeMeasurement} || \texttt{ReportData})$.
\end{enumerate}

\textit{We use generic terminology to abstract over vendor-specific components of a manufacturer's measured boot stack. In the Intel \gls{tdx} instantiation, \texttt{LaunchMeasurement} and \texttt{RuntimeMeasurement} correspond to \gls{mrtd} and \gls{rtmr}, respectively.}

On installation of an enclave, the $\mathcal{G}_{TEE}$ functionality spawns a new Interactive \gls{iti} subroutine with composite extended identity $(sh_{TEE}[\msf{prog}], (eid || pid, \msf{att} || idx))$:

\begin{shell}[label=func:sh_tdx, title={$sh_{TEE}[\msf{prog}]$}]

\heading{Parameters.}
Program $\msf{prog}$.

\heading{Interfaces.}
$\mathbb{O} = \mathbb{O}^{std}$,
$\mathbb{A} = \emptyset$.

\heading{Extended Identity.}
$(sh_{TEE}[\msf{prog}], (eid || pid, \msf{att} || idx))$.

\Init
\begin{itemizeless}
  \item Initialize an empty set of virtual ITIs.
\end{itemizeless}

\OnInput \inmsg{\msf{INSTALL}} from $\mathcal{G}_{att}^{mod}$:
\begin{itemizeless}
  \item \If virtual ITI $(\msf{prog}, (eid, idx))$ does not exist:
        \begin{itemizeless}
          \item Create virtual ITI $(\msf{prog}, (eid, idx))$.
        \end{itemizeless}
  \item \textit{(No platform functionality is required for the vanilla shell.)}
\end{itemizeless}

\OnInput \inmsg{\msf{inp}} from $\mathcal{G}_{att}^{mod}$:
\begin{itemizeless}
  \item Execute program $(\msf{prog}, (eid, idx))$ step by step on input $\msf{inp}$.
  \item For each instruction $i$:
        \begin{itemizeless}
          \item \If $i \in \mathbb{O}^{std}$:
                \begin{itemizeless}
                  \item Allow standard execution of $i$.
                \end{itemizeless}

          \item \Else \If $i = (\textbf{return } v)$:
                \begin{itemizeless}
                  \item \Output
                        $v$ with source
                        $(sh_{TEE}[\msf{prog}], (eid || pid, \msf{att} || idx))$.
                \end{itemizeless}
        \end{itemizeless}
\end{itemizeless}

\end{shell}

The behavior of the instantiated functionality $\mathcal{G}_{TEE}$ effectively replaces the generic "Install/Resume" logic of the shell with the hardware-specific operations of Intel \gls{tee}. This will later allow us to invoke AGATE's Theorem 6, which states that \textit{if a protocol is secure using the modular shell, it is secure under \gls{uc}}.

\subsection{The Global TPM Functionality ($\mathcal{G}_{TPM}$)}

We represent both discrete and virtual \glspl{tpm} using a Global Shared Resource functionality that provides stateful signing and sealing capabilities. The only distinction is the entity certifying the keys stored in the functionality's Registry $\mathcal{R}_{TPM}$: a discrete physical TPM's registry contains \glspl{ek} certified by the \gls{tpm} Manufacturer, whereas a \gls{vtpm} contains \glspl{ek} certified by the Cloud Provider. The security difference lies strictly in who populates $\mathcal{R}_{TPM}$ and the PCR Policy parameter: for discrete \glspl{tpm} in the bare metal scenario, this Policy includes the Intel \gls{txt} launch measurements (\glspl{pcr} 17-18).


\begin{functionality}[label=func:g_tpm, title=$\mathcal{G}_{TPM}(\mathcal{R}_{TPM})$]

\Init
\begin{itemizeless} 
  \item Initialize registry $\mathcal{R}_{TPM}$ with valid Endorsement Keys (EKs).
  \item Initialize state maps $\msf{PCRs}[\cdot]$ and $\msf{KeyStore}[\cdot]$.
\end{itemizeless}

\OnInput \inmsg{\msf{TPM2\_Create}, \msf{Policy}} from Party $P$:
\begin{itemizeless}
  \item \textbf{Leak:}
  Send \inmsg{\msf{CreateReq}, P} to the Adversary.
  Wait for \inmsg{\msf{Proceed}}.

  \item \textbf{Logic:}
  \begin{itemizeless}
    \item Generate key pair $(AK_{pub}, AK_{priv})$.
    \item Generate handle $h_{AK}$.
    \item \textbf{Bind Policy:}
          Store $(AK_{priv}, \msf{Policy})$ in $\msf{KeyStore}[h_{AK}]$.
    \item Return $(h_{AK}, AK_{pub})$.
  \end{itemizeless}
\end{itemizeless}

\OnInput \inmsg{\msf{TPM2\_PCR\_Extend}, \msf{index}, \msf{val}} from Party $P$:
\begin{itemizeless}
  \item \textbf{Leak:}
  Send \inmsg{\msf{ExtendReq}, P} to the Adversary.
  Wait for \inmsg{\msf{Proceed}}.

  \item \textbf{Logic:}
  \begin{itemizeless}
    \item Update
          $\msf{PCRs}[P][\msf{index}] \leftarrow
          \text{SHA384}(\msf{PCRs}[P][\msf{index}] || \msf{val})$.
    \item Return \msf{Success}.
  \end{itemizeless}
\end{itemizeless}

\OnInput \inmsg{\msf{TPM2\_Quote}, \msf{h_{AK}}, \msf{nonce}} from Party $P$:
\begin{itemizeless}
  \item \textbf{Leak:}
  Send \inmsg{\msf{QuoteReq}, P} to the Adversary.
  Wait for \inmsg{\msf{Proceed}}.

  \item \textbf{Logic:}
  \begin{itemizeless}
    \item Retrieve $(AK_{priv}, \msf{Policy})$ from $\msf{KeyStore}[h_{AK}]$.

    \item \textbf{Access Control (Sealing Check):}
    \begin{itemizeless}
      \item Check whether $\msf{PCRs}[P]$ satisfies $\msf{Policy}$.
      \item \textbf{If not:} Return \msf{Error} (simulating hardware refusal).
    \end{itemizeless}

    \item \textbf{Sign:}
    \begin{itemizeless}
      \item Construct payload
            $D \leftarrow (\msf{PCRs}[P] || \msf{nonce})$.
      \item Compute
            $\sigma \leftarrow \msf{Sign}_{AK_{priv}}(D)$.
    \end{itemizeless}

    \item Return $\sigma$.
  \end{itemizeless}
\end{itemizeless}

\end{functionality}

\paragraph{Modelling choices:} The \texttt{TPM2\_Create} command captures the creation of a restricted key. The \texttt{TPM2\_Quote} command enforces the restriction. This formally models the "Chain of Trust," in which a compromised Host (wrong \glspl{pcr}) cannot generate a valid quote. Defining $\mathcal{G}_{TPM}$ as a Global Functionality ensures that the \gls{pcr} state persists across different protocol invocations. This will be useful in the proof, allowing the simulator to catch the adversary if they fail to perform the correct measurements during boot.

\subsection{Ideal Functionality: $\mathcal{G}_{CloudTEE}$}

We define an alternative to $\mathcal{G}_{TEE}$ that has the extra ability to attest to its location in the cloud. Observe how it only differs from $G_{TEE}$ by having an additional \texttt{GetLocationProof} feature oracle. The $\mathcal{G}_{CloudTEE}$ functionality is defined as the modular instantiation:
$$\mathcal{G}_{CloudTEE} := \mathcal{G}_{att}^{mod}\left[\lambda, \mathcal{R}_{TEE}, \mathbb{O}_{CloudTEE}, \mathbb{A}_{TEE}, \mathbb{S}_{TEE}\right]$$

Where $\mathbb{O}_{CloudTEE} = \mathbb{O}_{TEE} \cup \{GetLocationProof\}$.

On installation of an enclave, the $\mathcal{G}_{CloudTEE}$ functionality spawns a new Interactive \gls{iti} subroutine with composite extended identity $(sh_{CloudTEE}[\msf{prog}], (eid || pid, \msf{att} || idx))$:


\begin{shell}[label=func:sh_cloudtdx_app, title={$sh_{CloudTEE}[\msf{prog}]$}]

\heading{Parameters.}
Program $\msf{prog}$.

\heading{Interfaces.}
$\mathbb{O} = \mathbb{O}^{std} \cup \{\msf{Quote}, \msf{GetLocationProof}\}$,
$\mathbb{A} = \emptyset$.

\heading{Extended Identity.}
$(sh_{CloudTEE}[\msf{prog}], (eid || pid, \msf{att} || idx))$.

\Init
\begin{itemizeless}
  \item Initialize an empty set of virtual ITIs.
\end{itemizeless}

\OnInput \inmsg{\msf{INSTALL}} from $\mathcal{G}_{att}^{mod}$:
\begin{itemizeless}
  \item \If virtual ITI $(\msf{prog}, (eid, idx))$ does not exist:
        \begin{itemizeless}
          \item Create virtual ITI $(\msf{prog}, (eid, idx))$.
        \end{itemizeless}
  \item \If ideal functionality $(\mathcal{G}_{TEE}, (idx, \bot))$ does not exist:
        \begin{itemizeless}
          \item Create ideal functionality $(\mathcal{G}_{TEE}, (idx, \bot))$.
        \end{itemizeless}
  \item \textit{(The shell ensures the existence of the helper functionality it relies on.)}
\end{itemizeless}

\OnInput \inmsg{\msf{inp}} from $\mathcal{G}_{att}^{mod}$:
\begin{itemizeless}
  \item Execute program $(\msf{prog}, (eid, idx))$ step by step on input $\msf{inp}$.
  \item For each instruction $i$:
        \begin{itemizeless}
          \item \If $i \in \mathbb{O}^{std}$:
                \begin{itemizeless}
                  \item Allow standard execution of $i$.
                \end{itemizeless}

          \item \Else \If $i = \msf{GetLocationProof}(\msf{nonce})$:
                \begin{itemizeless}
                  \item \Send \inmsg{\msf{LOC\_PROOF}, \msf{nonce}}
                        to $\mathcal{G}_{TEE}$.
                  \item Receive response $\lambda$.
                  \item Write $\lambda$ to the subroutine output of
                        $(\msf{prog}, (eid, idx))$.
                \end{itemizeless}

          \item \Else \If $i = (\textbf{return } v)$:
                \begin{itemizeless}
                  \item \Output
                        $v$ with source
                        $(sh_{CloudTEE}[\msf{prog}], (eid || pid, \msf{att} || idx))$.
                \end{itemizeless}
        \end{itemizeless}
\end{itemizeless}

\end{shell}

\subsection{The DCEA Protocol ($Prot_{DCEA}$)}
\label{sec:dcea_wrapper_protocol}

This is the internal shell for the wrapper protocol that only uses the standard \gls{tee} Quote oracle and implements the binding logic internally.

Note that the $\mathcal{G}_{att}^{mod}$ framework expresses calls to external functionalities as messages, hence why we do not directly call $\mathcal{G}_{TPM}$ but instead send a request to the Host, who then calls $\mathcal{G}_{TPM}$. The wrapper shell $W^{DCEA}$ is thus agnostic to the underlying root of trust protecting $\mathcal{G}_{TPM}$, and assumes that if a TPM quote verifies under $AK_{pub}$, then, by the guarantees of $\mathcal{G}_{TPM}$, that $AK_{pub}$ is bound to a specific platform state. Concretely, the specific protocols that hypothetically realize $\mathcal{G}_{TPM}$ would both provide a \gls{vtpm} interface as described in \Cref{sec:protocol-impl} of the main paper. The distinction in Root of Trust is determined by the scenario under consideration:

\begin{itemize}
	\item  In \textbf{Scenario I} (S1 - Managed CVM): The \gls{vtpm} is a software service provided by the cloud. The Root of Trust is the Cloud Provider's \gls{ca}.
  \item In \textbf{Scenario II} (S2 - Bare Metal): The \gls{vtpm} is a guest process running on the host. Its identity ($AK_{vTPM}$) is protected by a hardware-backed \gls{drtm} mechanism like Intel \gls{txt} or similar that extends measurements to \gls{pcr} registers. The physical TPM enforces that this $AK$ is only available if the host hypervisor and \gls{vtpm} binary match the measurements in designated \gls{txt} \glspl{pcr} 17-18. 
\end{itemize}

In both cases, $W^{DCEA}$ remains identical: it binds its execution to the vTPM it indirectly accesses. The Proof handles the distinction of whether that \gls{vtpm} is anchored by a Provider policy (\textbf{S1}) or Hardware Sealing (\textbf{S2}).

\begin{shell}[label=func:w_dcea, title={$W^{DCEA}[\msf{prog}]$}]

\heading{Parameters.}
Program $\msf{prog}$.

\heading{Identity.}
$(eid || c, idx)$.

\heading{Parent Shell.}
$(sh_{TEE}[W^{DCEA}[\msf{prog}]], (eid || pid, \msf{att} || idx))$.

\heading{Oracles Used.}
$\mathbb{O} = \mathbb{O}^{std} \cup \{\msf{Quote}\}$.

\Init
\begin{itemizeless}
  \item Initialize an empty set of virtual ITIs.
\end{itemizeless}

\OnInput \inmsg{\msf{INIT}} from $(eid || pid, \msf{att} || idx)$:
\begin{itemizeless}
  \item Install virtual ITI
        $(\msf{prog}, (eid || c || \msf{wrapped}, idx))$.
\end{itemizeless}

\OnInput \inmsg{\msf{inp}} from $(eid || pid, \msf{att} || idx)$:
\begin{itemizeless}
  \item Execute program
        $(\msf{prog}, (eid || c || \msf{wrapped}, idx))$
        step by step on input $\msf{inp}$.
  \item For each instruction $i$:
        \begin{itemizeless}

          \item \If $i = \msf{GetLocationProof}(n_V)$:
                \begin{itemizeless}
                  \item \Send
                        \inmsg{\msf{RESUMEREQUEST},
                        (\msf{TPM\_QUOTE\_REQ}, $n_V$)}
                        to the parent shell.
                  \item Await response from the Host.
                  \item \If the next message is
                        \inmsg{\msf{TPM\_QUOTE\_RESP},
                        $\sigma_{vTPM}, AK_{pub}$}:
                        \begin{itemizeless}
                          \item Compute binding blob $B \leftarrow \text{SHA384}(AK_{pub} || n_V)$.
                          \item Request TD quote embedding $B$: $\sigma_{TEE} \leftarrow \msf{\mathcal{G}_{TEE}.Quote}(B)$.
                          \item Construct composite proof $\Pi \leftarrow(\sigma_{vTPM}, \sigma_{TEE}, AK_{pub}).$
                          \item Write $\Pi$ to the subroutine output of
                                $(\msf{prog}, (eid || c || \msf{wrapped}, idx))$.
                        \end{itemizeless}
                  \item \Else abort execution.
                \end{itemizeless}

          \item \Else
                \begin{itemizeless}
                  \item Allow standard execution of instruction $i$.
                \end{itemizeless}
        \end{itemizeless}
\end{itemizeless}

\end{shell}

\subsection{Proof sketch}

\begin{theorem}
Let $G_{TEE}$ denote the vanilla instance $G_{att}^{mod}[\dots, sh_{TEE}]$ and
let $G_{CloudTEE}$ denote the target instance $G_{att}^{mod}[\dots, sh_{CloudTEE}]$.
Then the protocol $Prot_{DCEA}$, which installs the wrapper shell $W^{DCEA}$,
in the presence of $G_{TEE}$ UC-emulates $G_{CloudTEE}$ against any static adversary
$\mathcal{A}$ controlling the Host.
\end{theorem}

\begin{proof}
The proof proceeds via the AGATE composition framework. We anchor the real-world protocol description to Intel \gls{tdx} and \gls{txt} terminology when necessary to facilitate comprehension.

\paragraph{Step 1: Reduction via AGATE composition.}
We invoke \textit{Theorem 6} (General Replacement of Global Setups) from the AGATE framework.
This theorem states that if a wrapper protocol $\mathcal{W}$ combined with a weaker global setup $G_{att}$ UC-emulates a stronger setup $G_{att}'$, then $\mathcal{W}$ can replace the stronger setup in any larger protocol context.

In our setting, $G_{TEE}$ serves as the weaker setup and $G_{CloudTEE}$ as the stronger setup. It therefore suffices to show that the wrapper shell $W^{DCEA}$ correctly emulates the behavior of the ideal shell $sh_{CloudTEE}$, in particular for the instruction \texttt{GetLocationProof}, using only the vanilla \texttt{Quote} oracle and interaction with the host-provided vTPM.

Although \gls{vtpm} quotes are generated by $\mathcal{G}_{TPM}$ in the real execution, the wrapper receives them via the host-controlled interface. Still, this does not impact quote validity. Any quote verifying under $AK_{pub}$ must originate from a \gls{tpm} whose endorsement key is certified and whose attestation key is bound to the measured platform state. In \textbf{Scenario I}, this binding is backed by the cloud provider. In \textbf{Scenario II}, the binding is enforced via Intel \gls{txt} measurements extended into \glspl{pcr}. The distinction between \textbf{Scenario I} and \textbf{II} and the corresponding roots of trust was discussed in Section \ref{sec:dcea_wrapper_protocol}.

\paragraph{Step 2: Simulator construction.}
We construct a simulator $\mathcal{S}$ that interacts with the environment
$\mathcal{Z}$, the ideal functionality $G_{CloudTEE}$, and the real-world adversary
$\mathcal{A}$ controlling the Host.

\subparagraph{Setup and signature transformation.}
In the ideal world, the guest program is executed under $sh_{CloudTEE}$.
In the real world, it is executed under $sh_{TEE}$ running the wrapped program
$W^{DCEA}[\textit{prog}]$.
To make these executions indistinguishable, $\mathcal{S}$ applies a signature
transformation $F$ as defined in AGATE Conjecture 2.
Whenever $G_{CloudTEE}$ outputs a signature on $\textit{prog}$, $\mathcal{S}$
transforms it into a signature that is consistent with the execution of
$W^{DCEA}[\textit{prog}]$, thereby hiding the presence of the wrapper from $\mathcal{Z}$.

\subparagraph{Modelling the Roots of Trust} The simulator initialises an internal $\mathcal{G}_{TPM}$ instance based on the corrupted party set and deployment scenario: $\mathcal{S}$ populates $\mathcal{R}_{TPM}$ with Cloud Provider-signed keys under \textbf{Scenario I} (\textbf{S1}) and with Manufacturer-signed keys under \textbf{Scenario II} (\textbf{S2}). Crucially, $\mathcal{S}$ must enforce Intel \gls{txt}'s security properties: it maintains a simulated \gls{pcr} state for the physical \gls{tpm} so that it can update the simulated \gls{pcr} values according to the adversary $\mathcal{A}$'s behavior. Any attempt by $\mathcal{A}$ to unseal or use the legitimate $AK_{vTPM}$, or to modify the expected host OS or vTPM binary, will fail within the $\mathcal{G}_{TPM}$ simulation, correctly reflecting hardware sealing in the real world.

\subparagraph{Simulation of execution.}

When the Adversary $\mathcal{A}$ tries to interact with the \gls{vtpm} or the \gls{tpm}, $\mathcal{S}$ captures these calls and services them using its internal $G_{TPM}$ simulation. $\mathcal{S}$ records all valid keys $\{AK_i\}$ and quotes $\{\sigma_i\}$ generated by $\mathcal{A}$.

Upon receiving a \texttt{RESUME} command from $\mathcal{Z}$, the simulator executes
the code of $W^{DCEA}[\textit{prog}]$ internally.

If the next instruction is a standard instruction, $\mathcal{S}$ forwards execution
to $G_{CloudTEE}$.

If the instruction is \texttt{GetLocationProof}, $\mathcal{S}$ simulates the wrapper
logic by sending a simulated
\[
(\texttt{RESUMEREQUEST}, (\texttt{TPM\_QUOTE\_REQ}, n))
\]
message to the adversary $\mathcal{A}$ and receiving a \gls{vtpm} quote
$\sigma_{vTPM}$ together with the public key $AK_{pub}$.

\paragraph{Step 3: Handling the proxy attack.} If $\mathcal{A}$ returns a \gls{vtpm} quote originating from a different platform than the simulated execution instance, $\mathcal{S}$ detects the inconsistency. In \textbf{Scenario II}, such inconsistencies cannot be masked by replaying quotes from another machine, because the attestation key is sealed to \gls{txt}-extended \gls{pcr} values and cannot be used off-platform. The simulator computes the same binding value
\[
B = \text{SHA384}(AK_{pub} \parallel n)
\]
that the real wrapper would compute.

If the \gls{vtpm} evidence is consistent with the simulated execution, $\mathcal{S}$
allows $G_{CloudTEE}$ to proceed and reformats the ideal output into a composite proof matching the real-world output format.

If the evidence is inconsistent (the proxy scenario), then in the real world, the wrapper produces a proof that necessarily fails verification. The Real World wrapper relies on the \gls{vtpm}'s $AK$. Depending on the scenario, the Cloud Provider or Intel \gls{txt} controls access to the signing $AK$. Any attempt from $\mathcal{A}$ to proxy the request to a different machine will either lack the sealed \gls{vtpm} blob or fail to unseal it due to \gls{pcr} mismatches. Therefore, $\mathcal{A}$ cannot produce a valid \gls{vtpm} quote from a remote machine. $\mathcal{S}$ observes that no valid quote can be produced and mirrors $G_{CloudTEE}$'s behavior by instructing $G_{CloudTEE}$ to abort or output a value that fails verification, thus preserving consistency between the two worlds.

\paragraph{Step 4: Indistinguishability.}
We argue that no environment $\mathcal{Z}$ can distinguish the real and ideal
executions. Indistinguishability follows from the unforgeability of hardware-backed signatures and the correctness of the signature transformation. We conclude that $Prot_{DCEA}$ UC-emulates $G_{CloudTEE}$.
\end{proof}